\documentstyle[11pt]{article}

\begin{document}
\begin{titlepage}
\rightline{IC/94/399}
\rightline{SISSA-181/94/EP}
{}~\\
\begin{center}
{\Large \bf Neutrino spin-flip effects in collapsing stars}
\vspace{0.4cm}
\end{center}
\begin{center}
{\large H. Athar}$^{(1,4)}$,
{\large J. T. Peltoniemi}$^{(2)}$, and
{\large A.Yu. Smirnov}$^{(1,3)}$
\\
\vspace{0.2cm}
(1){\em International Centre for Theoretical Physics, I-34100 Trieste, Italy}\\
(2){\em INFN, Trieste and International School for Advanced Study \\
I-34013 Trieste, Italy} \\
(3){\em Institute
for Nuclear Research, Russian Academy of Sciences, Moscow, Russia}\\
(4){\em Department of Physics,
Quaid-i-Azam University, Islamabad, Pakistan}

\end{center}

\begin{abstract}
We study the spin-flavor transitions of neutrinos,
$\bar{\nu}_e - \nu_{\mu}$,
$\nu_e - \bar{\nu}_{\mu}$, etc., in the
magnetic fields of a collapsing star.
For the neutrino mass squared difference
$\Delta m^2 \sim (10^{-10} - 10)$ eV$^2$ the transitions take place
in almost isotopically neutral region of the star,
where the effective matter density
is suppressed up to  3 -- 4 orders of magnitude.
This suppression is shown to increase the
sensitivity of the neutrino bursts studies to the
magnetic moment of neutrino, $\mu$,
by 1.5 -- 2 orders of magnitude, and for realistic
magnetic field the observable effects may exist for $\mu \sim
(2 - 3)\cdot 10^{-14} \mu_B$ ($\mu_B$ is the Bohr magneton).
In the isotopically neutral region the jumps of the effective
potential exist which influence the probabilities of transitions.
The experimental signatures of the spin-flavor transitions
are discussed.  In particular, in the case of direct mass hierarchy,
the spin-flip effects result in a variety of modifications of the
$\bar{\nu}_e$-spectrum. Taking this into account, we estimated the upper
bounds on $\mu B$ from the SN1987A data.
In the isotopically neutral region the effects of
possible twist of the magnetic field on the way of neutrinos can be
important, inducing  distortion of the neutrino energy spectra and
further increasing the sensitivity to $\mu B$. However, if the
total rotation angle is restricted by $\Delta \phi < \pi$,
the absolute change of probabilities is small.

\end{abstract}
\end{titlepage}
\section{Introduction}

At least 100 neutrino bursts from the gravitational collapses of stars in our
Galaxy are already on the way to the Earth\footnote{This comes about
from the estimation of frequency of the collapses in our Galaxy as
1/30 year \cite{van}}. A registration of
even one of those by the existing detectors, Kamiokande \cite{kam}, Baksan
\cite{baks}, LSD \cite{lsd}, and especially, by the new installations:
LVD \cite{lvd} (which is  already working),
SNO \cite{sno} and  Superkamiokande \cite{skam}
(starting to operate in 1996),  ICARUS \cite{icar} (which is at the
prototype stage) will give unique and extremely rich information about
features of
the gravitational collapse, supernova phenomena and neutrinos
themselves. Already SN1987A  has given a lot.
In general, one will be able to  get information on the energy
spectra (as a function of time) of $\nu_e$,  $\bar{\nu}_e$,
as well as of neutrinos of the non-electron type:
$\nu_{\mu}$ $\bar{\nu}_{\mu}$, $\nu_{\tau}$,  $\bar{\nu}_{\tau}$.
Certain properties  of these spectra
do not depend on the model of the star thus opening the
possibility to study the characteristics of neutrinos.
New experiments are discussed
to detect the neutrino bursts from collapses in other
nearby galaxies \cite{DBC}.

In this paper we  consider a possibility to study  effects of
the neutrino magnetic moments using  the neutrino bursts.
In the magnetic fields the neutrinos
with  magnetic moments undergo  spin-precession \cite{AC,KF,MBV},
spin-flavor precession \cite{JS} or/and resonant spin-flavor conversion,
e.g. $\nu_{eL}\rightarrow \bar{\nu}_{\mu R}$ \cite{lim,Ekh}.
Previously, the applications of these
effects to neutrinos from supernova have been considered in \cite{KF}
(precession) and \cite{lim,AB} (resonance conversion).

The spin-flip effects are  determined by the product, $\mu B$,
of the neutrino magnetic moment, $\mu$, and the
strength of the magnetic field, $B$.
The discovery of the effects
corresponding to values of  $\mu$ near the present upper bound,
$\mu < (1  - 3)\cdot 10^{-12}\mu_B$ \cite{mom}
($\mu_{B}$ is the Bohr magneton)
will imply rich   physics beyond the standard model.
{}From this point of view the studies of the neutrino bursts are of special
interest since the magnetic fields in supernova can be as strong as
$(10^{12} - 10^{14})$ Gauss.

It is
reasonable to address  two questions, keeping in mind that the
magnetic field profiles of supernovae  are essentially unknown,
and vary  from the star to star.
What are the experimental
signatures of the spin-flip effects? And what could be the sensitivity
of the neutrino burst studies to the magnetic moments of neutrinos,
or more precisely to $\mu B$?

The spin-flip probability may be of the order $1$, if
\begin{equation}
 \mu  {\buildrel > \over {_{\sim}}}    \frac{1}{\int {\rm d}r B(r)}~,
\end{equation}
where the strength of the magnetic field
 is integrated along neutrino trajectory.
Suggesting the field profile
\begin{equation}
 B \simeq B_{0}\left(\frac{r_{0}}{r}\right)^{k},
\end{equation}
with $B_{0} \sim  (10^{12}-10^{14})$ Gauss, $k = 2-3$,
and $r_{0}\sim 10$ km,
where $r$ is the distance from the center of star,
one gets from  (1),
\begin{equation}
\mu B_0 {\buildrel > \over {_{\sim}}}    \frac{k-1}{r_0}
\end{equation}
and  $\mu  {\buildrel > \over {_{\sim}}}    (10^{-15}-10^{-17}) \mu_{B}$.
(For a constant field a similar bound is
$\mu B \geq (\Delta r_B)^{-1}$, where $\Delta r_B$ is
the size of region with the magnetic field).
These numbers look encouraging, being  3 - 5 orders
of magnitude below the present bound.

However,  the estimation (1)
corresponds to neutrino propagation in vacuum (no matter),
where the precession
takes place with maximal depth, $A_P = 1$. In this case  the ``optimal
conditions" for the spin-flip are realized:
the  value of $\mu B$ needed to change the spin-flip probability
by a given amount $\Delta P$ is minimal \cite{M}.
The presence of dense matter strongly reduces
a sensitivity to $\mu$. Forward neutrino scattering (refraction) results
in neutrino level splitting: $\Delta H =  V_{SF}$, where $V_{SF}$ is
the difference of the effective potentials acquired by the left and right
handed neutrino components in matter. Typically $V_{SF} \sim V_0$, where
\begin{equation}
 V_0  \equiv
\sqrt{2}G_{F}n \approx \sqrt{2}G_{F}\rho/m_N
\end{equation}
is the total potential, $G_{F}$ is the Fermi
constant, $n$ is the nucleon number density,
$\rho$ is the matter density and $m_N$ is the nucleon mass.
The level splitting suppresses the
depth of precession \cite{MBV}:
\begin{equation}
A_{P} =  \frac{(2\mu B)^2}
{(2\mu B)^2 +  (\Delta H)^2}
= \frac{1}{1+\left[V_{SF}/2\mu B\right]^{2}},
\end{equation}
and to have   $A_{P}\, \geq \, \frac{1}{2}$, one needs
\begin{equation}
\mu B  \geq \frac{1}{2} V_{SF}.
\end{equation}
In dense medium the restriction (6) is much stronger than (3).
Indeed, for typical density of the neutrinosphere of the supernova:
$\rho \sim  10^{12}$ g/cm$^{3}$ and
for $B\,\simeq \, 10^{14}$ Gauss, one gets
from (6): $\mu  \geq  10^{-7} \mu_{B}$.  With
increase of distance, the potential $V_{SF}$ decreases as
$n \propto  r^{-3}$, or quicker, and
the restriction (6) relaxes faster than (3).
At $r {\buildrel > \over {_{\sim}}}    R_{\odot}$
it becomes even weaker  than (3)
(here $R_{\odot} \equiv  7\cdot 10^{10}$ cm
is the solar radius).

The  suppression of the matter density (i.e., approaching the vacuum
condition) results in the increase of the sensitivity to $\mu$. This is
realized
in the case of the resonant spin-flavor conversion \cite{Ekh}. If the left and
right handed
neutrino components connected by the magnetic moment have different masses,
$m_1$, $m_2$, being also of different flavors, the matter effect can
be compensated by mass splitting:
\begin{equation}
\Delta H = V_{SF} - \frac{\Delta m^{2}}{2E} = 0
\end{equation}
({\em resonance condition}), where $\Delta m^2 \equiv m_2^2 - m_1^2$,
and $E$ is the neutrino energy.
Now the expression for $A_{P}$ becomes
\begin{equation}
 A_{P} = \frac{(2\mu B)^{2}}{(2\mu B)^{2}+\left(V_{SF} -\frac{
 \Delta m^{2}}{2E}\right)^{2}},
\end{equation}
and in the resonance (7) one gets $A_{P}\,=\,1$.
However, since  $V_{SF}$ changes with distance, the equality  $A_{P} = 1$
holds only in the resonance point, $r_{R}$.
Strong  compensation of the matter effects occurs in some
layer  around $r_{R}$ whose size
depends on  gradient of the potential:
$\dot{V}_{SF} \equiv  {\rm d}V_{SF}/{\rm d}r$. Indeed,
from (8) one finds that $A_{P} \geq \frac{1}{2}$, if
$\Delta H = \Delta V_{SF} \leq \Delta V_R$, where
\begin{equation}
\Delta V_R =  2\mu B
\end{equation}
is the half-width of the resonance layer. The
spatial size of the region with $A_P \geq 1/2$ equals
$2 \Delta r_{R} = 2 (\dot{V}_{SF})^{-1}\Delta V_R
= 2 (\dot{V}_{SF})^{-1} 2\mu B$. Then the strong spin-flip effect implies
that $2 \Delta r_R$ is larger than half of the precession length,
$l_p$: $2 \Delta r_R {\buildrel > \over {_{\sim}}}    l_p/2 = \pi/2 \mu B$.
Substituting the expression for
$\Delta r_R$
in this inequality
one gets
\begin{equation}
\kappa_R \equiv  \frac{2 (2\mu B)^{2}}{\pi |\dot{V}_{SF}|}
{\buildrel > \over {_{\sim}}}    1.
\end{equation}
This  is the {\em adiabaticity condition} in the resonance \cite{lim,Ekh}, and
$\kappa_R$ is the resonance adiabaticity parameter.
(Note that the adiabaticity parameter $\gamma$ used in \cite{Ekh} is related
to that in (10) as  $\kappa_R = 2\gamma/\pi$).
For large densities the
condition (10) is much less restrictive, than (6),  allowing for a
strong spin-flip effect for smaller values of $\mu B$. For example, at $B =
10^{14}$ Gauss, $\rho = 10^{12}$ g/cm$^3$,  and $V_{SF}/\dot{V}_{SF}
\sim  r_0  \sim 10$ km, one gets $\mu  > 3\cdot 10^{-12} \mu_{B}$,
instead of $10^{-7} \mu_B$.

Here we will consider {\it two new phenomena} which suppress the
matter effect and thus, increase the
sensitivity of neutrino burst studies to the neutrino magnetic
moment.
The potential $V_{SF}$ can be diminished for  special nuclear composition of
matter.  We point out that this is realized in the external region
of supernova, where the medium is  almost isotopically
neutral. Another phenomenon  is the  magnetic field twist --
the change of the direction of the magnetic strength lines on the way of
neutrinos. The field twist  leads to
an additional contribution to the level splitting $\Delta H$
which  can compensate the matter effect.

The experimental signatures of the spin-flavor transitions of supernova
neutrinos
have been
discussed previously in  \cite{AB}. The influence of the spin-flip
on the relation between directional (induced by $\nu_e e$-scattering) and
isotropic ($\bar{\nu}_e p$ - interaction)
signals was studied. Without consideration of the
dynamics of the propagation it was suggested that the  probabilities
of different transitions are either
one  or zero. The estimated number of
events shows that four cases (1) no transitions, (2) only flavor
transitions, (3) only spin-flavor transitions,
(4) flavor plus spin-flavor transitions can be disentangled by
new installations (SNO, Superkamiokande). We will consider a
more general situation concentrating on the energy dependence of
the transitions.

In this paper we study  the dynamics of propagation to estimate
the values of $\mu B$ needed for different effects. We  consider
signatures of the spin-flip transitions related to specific density
distribution
in the isotopically neutral region and to possible presence of the
magnetic field twist. The paper is organized as follows.
In Sect.~2, we describe the effective potential in supernova.
In Sect.~3 the level crossing schemes for
different values of neutrino masses are found and the dynamics
of neutrino propagation is considered. In Sect.~4, we
study the dependence of the transition probabilities on
the magnetic field profile and on the neutrino parameters.
Sect.~5 is devoted to the interplay of the spin-flavor and flavor
transitions.  In Sect.~6  we consider some special  effects of
neutrino propagation and discuss the possible implications of the results.
In particular, we describe the distortion of
energy spectra of neutrinos and estimate the sensitivity of the studies of
$\nu$-bursts to $\mu B$. Also, the upper bound on ($\mu B$) will be
obtained from SN1987A data. In Sect.~7, the effects of the magnetic field
twist are considered.

\section{
Isotopically neutral region. Effective  potential.}

For the spin-flavor transitions,
e.g., $\nu_{eL}\rightarrow \bar{\nu}_{\mu R},\,\bar{\nu}_{eR}\rightarrow
\nu_{\mu L}$, the matter effect
is described by the potential \cite{lim,Ekh}
\begin{equation}
V_{SF} = \sqrt{2}G_{F}n(2Y_{e}-1) \equiv  V_0~ (2Y_{e}-1),
\end{equation}
where $Y_{e}$ is the number of electrons per nucleon. The value of $Y_e$
depends on  the nuclear composition of the matter. In isotopically neutral
medium (\# protons = \# neutrons),  one has  $Y_{e}\,=\,\frac{1}{2}$,
and according to (11)  $V_{SF} =  0$.
Consequently, the matter effect is determined by the deviation
from the isotopical neutrality.

For the flavor conversion the potential is proportional to the electron
density:
\begin{equation}
V_{F}  \simeq  \sqrt{2}G_{F}nY_{e} = V_0 Y_{e}~~.
\end{equation}
Evidently, there is no suppression of $V_F$ in the isotopically neutral
medium; $V_F$ is suppressed in the central
strongly neutronized region of the star.

\subsection{Isotopically neutral region}

The progenitor of the  type II supernova has the ``onion"
structure. Below the hydrogen envelope the layers  follow
which consist mainly of
the isotopically neutral nuclei:
${}^{4}$He,  ${}^{12}$C, ${}^{16}$O, ${}^{28}$Si, ${}^{32}$S.
Thus the  region  between the hydrogen
envelope and the core,  where elements of iron peak dominate,
is almost isotopically neutral. The deviation
from the neutrality is stipulated by  small
abundances, $\xi_{i}$,  of the elements with small
excess of neutrons,
$i = {}^{22}$Ne, $^{23}$Na, ${}^{25}$Mg,  ${}^{56}$Fe \cite{ST,SNT}.
It can be written as
\begin{equation}
(1-2Y_{e}) \approx  \sum_{i}\xi_{i}\cdot
\left(1-\frac{2Z_{i}}{A_{i}}\right),
\end{equation}
where $Z_{i}$ and $A_{i}$ are the electric charge and the atomic
number of nuclei ``$i$", correspondingly.
Typically, one gets $\sum_{i} \xi_{i} \leq  10^{-2}$
and $(1-\frac{2 Z}{A}) \leq  10^{-1}$, and
therefore, according to (13): $(1-2Y_{e})\,\leq \, 10^{-3}$.
The value of $(2Y_{e}-1)$ equals
\begin{equation}
 (2Y_e - 1) = \left\{ \begin{array}{ll}
                        0.6-0.7 & \mbox{hydrogen envelope}  \\
                        -(10^{-4}-10^{-3}) & \mbox{isotopically neutral
                                           region}  \\
                        -(0.2-0.4) & \mbox{central regions of star}
\end{array}
              \right.
\end{equation}
Consequently, in the isotopically neutral region
the potential  $V_{SF}$  is suppressed by more
than 3 orders of magnitude with respect to the total potential
$V_0$ (fig.1). We  calculated the potentials  using the
model of the progenitor with mass  $15 M_{\odot}$
(where $M_{\odot}$ is the solar mass) \cite{ST}.
Two remarks are in order.
For $r < R_{\odot}$ the spatial distributions of the total density,
are very similar (up to factor 2) for stars in a
wide range of masses: $M =  (12-35) M_{\odot}$.
The  profiles may differ appreciably  in the external low density regions
which are  unessential for our consideration.
The collapse of the core and the propagation of the shock
wave  change both the
density profile and the nuclear composition of medium. However, during
the neutrino burst (0 - 10 seconds after the core bounce) the shock wave
can reach the distance $\sim$ several 1000 km at most. Therefore
most part of the
isotopically neutral region turns out to be undisturbed and
one can use the profile of the progenitor.

Strong convection effects in the inner and intermediate parts of the star,
if exist, may result in an injection of the elements in the iron peak into
outer layers, thus diminishing the degree of the isotopical neutrality.

\subsection{Properties of the effective potential}

According to (14) and Fig. 1, the effective
potential $V_{SF}$
has the following features:
It is positive and of the order $V_0$ in the H-envelope;
$V_{SF}$ decreases quickly and changes the sign in the border
between the hydrogen envelope and the ${}^{4}$He-layer ($r \sim  R_{\odot}$);
$V_{SF}$ is negative and suppressed by $3-4$ orders of magnitude in
comparison with $V_0$ in the isotopically neutral region;
$V_{SF}$  quickly increases in the inner edge of the
isotopically neutral region ($r \sim  10^{-3} R_{\odot}$),
and becomes again of the order $V_0$ in the central
part of the star. In the isotopically neutral region the jumps of the
effective potential $V_{SF}$ (not $V_0$!) exist which
are related to the change of the nuclear composition in the layers
of local nuclear ignition. The jump may be as
large as an order of magnitude.

The isotopically neutral region spreads
from $10^{-3}R_{\odot}$ to $\approx
R_{\odot}$ (Fig.~1). Here the potential changes
from $V_{SF} = (10^{-9} - 10^{-8})$ eV to zero:
$V_{SF} \rightarrow 0$, when $r \rightarrow (0.8 - 1)R_{\odot}$.
Below $10^{-17}$ eV, however,
$V_{SF}(r)$ decreases so quickly
that no matter effects are induced.
The  detectable energy range of the neutrinos from the gravitational
collapse is $(5 - 50)$ MeV. Consequently,
in the isotopically neutral region the resonance condition
(7) is fulfilled for
$\Delta m^{2} \sim (10^{-10}-0.1)$ eV$^{2}$.
This covers the range of $\Delta m^{2}$
interesting from the point of view of the solar and the atmospheric
neutrino  problems.
For the cosmologically interesting values, $\Delta m^2 \sim (4 - 50)$ eV$^2$,
the resonance is at the inner part of the isotopically neutral region,
where the potential is still suppressed by factor $1/30$ -- $1/100$.

Let us stress that the suppression of the effective density takes
place for the spin-flavor conversion only, i.e. when both neutrino components
are active. In  the  case of the spin-flip into sterile neutrino
$\nu_{e}\rightarrow \bar{\nu}_{s}$, the effective potential  equals
$V_s  =  \frac{V_0}{2}(3Y_{e}-1)$,
and in the isotopically neutral region one has
$V_s  = \frac{V_0}{4}$. The potential $V_s$  is suppressed in
strongly neutronized central part of star, where
$Y_{e}  \approx  \frac{1}{3}$ \cite{vol,pelt}.

\section{The dynamics of the neutrino transitions}

\subsection{Level crossing scheme}

We will consider the system of three massive neutrinos with
transition magnetic moments and vacuum mixing
(see \cite{APS} for details). For definiteness we assume the direct
mass hierarchy: $m_1 \ll m_2 \ll m_3$
and the smallness of flavor mixing.
The  energies of the flavor levels
(the diagonal elements of the effective Hamiltonian),
$H_{\alpha}$, $(\alpha  =  \nu_{e},  \bar{\nu}_{e},
\nu_{\mu}, \bar{\nu}_{\mu},
\nu_{\tau}, \bar{\nu}_{\tau})$, can be written as
\begin{equation}
\begin{array}{l}
H_{e} \equiv 0, \\
H_{\bar{e}} \equiv  -V_0~ (3Y_{e}-1)-\dot{\phi}, \\
H_{\mu} \equiv \cos 2\theta ~
\frac{\bigtriangleup m^{2}}{4E}  - V_0~ Y_{e},\\
H_{\bar{\mu}}  \equiv
\cos 2 \theta ~ \frac{\bigtriangleup m^{2}}{4E}
 - V_0~ (2Y_{e}-1)-\dot{\phi}.
\end{array}
\end{equation}
Here  $\theta$ is the $e$-$\mu$-mixing angle in vacuum.
The  $\nu_{\tau}$ energy levels,
$H_{\tau}$, $H_{\bar{\tau}}$, can be obtained from
$H_{\mu}$,   $H_{\bar{\mu}}$ by substituting $\theta \rightarrow
\theta_{\tau}$ and $\Delta m^2 \rightarrow \Delta m_{31}^2$.
The angle $\phi (t)$ ($\dot{\phi}\equiv {\rm d}\phi/{\rm d}r$)
defines the direction of the magnetic field in the transverse plane.
In (15) for later convenience, we have subtracted
from all elements the energy $H_e$
to make the first diagonal element of the Hamiltonian (i.e., H$_{e}$)
equal to zero.
(This is equivalent to the renormalisation of all wave functions by the same
factor
which does not change the probabilities).  The qualitative
dependence of the energy levels $H_{\alpha}$
on distance $r$ for $\dot{\phi} = 0$  and for
different values of $\Delta m^{2}/2E$
is shown in Fig.~2. The resonance (level crossing) conditions read as
$H_{\alpha} = H_{\beta}$ ($\alpha \neq \beta$).

The peculiar behavior  of the effective potential $V_{SF}$
in the isotopically neutral
region stipulates a number of features  in  the
level crossing scheme.
With increase of $\Delta m^2$ one gets the following changes.
(The levels for $\nu_e$ and $\bar{\nu}_e$ are fixed, whereas
$\nu_{\mu}$- and $\bar{\nu}_{\mu}$-levels go up).

For $\Delta m^{2} {\buildrel < \over {_{\sim}}}
10^{-10}$eV$^{2}$ ($E \sim 20$ MeV)
the mass splitting can be neglected.
Two  crossings of the levels  which correspond to the spin-flavor
conversions
$\nu_{e}\rightarrow \bar{\nu}_{\mu}$ and $\bar{\nu}_{e}\rightarrow
\nu_{\mu}$
occur practically at the same point,
where $(2Y_{e}-1) \simeq  0$ ($r \simeq  0.8R_{\odot}$), i.e.
on the border between the H-envelope and
${}^{4}$He-layer. The crossings are
stipulated by change of nuclear composition of the star (Fig. $2$a).

For $\Delta m^{2} {\buildrel > \over {_{\sim}}}    10^{-10}$ eV$^{2}$
the spin-flavor resonances, $\nu_{e}- \bar{\nu}_{\mu}$
and $\bar{\nu}_{e} - \nu_{\mu}$, are
spatially splitted. With increase of $\Delta
m^{2}$ the $\bar{\nu}_{e}-\nu_{\mu}$ resonance shifts
to the center of star, whereas $\nu_{e}-\bar{\nu}_{\mu}$
does to the surface.
Moreover, second resonance of ($\bar{\nu}_{e} - \nu_{\mu}$)-type appears
in the outer layers. When  $\Delta m^{2}$ increases
these same resonances approach each other, and at
$\Delta m^{2} \sim  3\cdot 10^{-8}$ eV$^{2}$ merge
(a ``touching" point of $\nu_{e}$- and $\bar{\nu}_{\mu}$-levels)
(Fig. 2b).

For $\Delta m^{2} {\buildrel > \over {_{\sim}}}
3\cdot 10^{-8}$ eV$^{2}$, there is only one
spin-flavor resonance, $\bar{\nu}_{e}-\nu_{\mu}$. The flavor resonance
$\nu_{e}-\nu_{\mu}$ lies  closer to the surface.  When
$\Delta m^{2}$
increases both these resonances shift to the center (Fig. 2c).

The  $\nu_{\tau}$-level can cross $\nu_e$- and $\bar{\nu}_e$-levels in the
isotopically neutral region. Moreover,
for cosmologically interesting values of $\Delta m^2_{13}$
the $\nu_{\tau}$-resonances are in
inner edge of this region. They can  also
be in the central part of the star.

\subsection{Adiabaticity conditions}

In the isotopically neutral region
the effective potentials for the flavor conversion and the spin-flavor
conversion differ by $3-4$ orders of magnitude.
Therefore,  the flavor  and spin-flavor
resonances are strongly separated in space, in
contrast with the case of usual medium.
Consequently,  one can consider these crossings independently,
reducing the task to two neutrino tasks.

If  $\theta$ is small
the flavor transition is essentially a local resonance  phenomenon.
It takes place at $r_R$ determined from (15): $H_{\mu}(r_R) = 0$.
Therefore, the
effect depends on the local properties of the density distribution
in resonance:
$V_{F}(r_{R}) = V_0 Y_e(r)$ and $ H_{F}  \equiv  V_{F}/\dot{V}_{F}|_{r_{R}}$.
The flavor adiabaticity reads
\begin{equation}
\kappa_R^F \equiv  \frac{H_{F}}{\pi} \cdot \frac{\Delta m^{2}}{E}\cdot
\frac{\sin^{2}2\theta}{\cos
 2\theta} > 1.
\end{equation}
(This corresponds to the situation when at least half of the oscillation
length,
is obtained in the resonance
layer,
in the direct analogy with the definition of the adiabaticity parameter
for the spin-flip conversion (10)).
Using the potential $V_{F}\sim V_0$  (fig.~1), we find that for neutrinos
with
$E \sim 20$ MeV and $\Delta m^{2} = 10^{-6}, 10^{-5}, 10^{-4}$ and
$10^{-2}$ eV$^2$, the resonance is situated
at $r =
0.7 R_{\odot}$, $0.5 R_{\odot}$, $0.17 R_{\odot}$,
$0.03 R_{\odot}$, correspondingly,
and the condition
(16) is satisfied for  $\sin^{2}2\theta  > 0.5,~ 0.06,~ 0.02$, $10^{-3}$.

In contrast, the  spin-flip effects can be nonlocal even in case of variable
density. Indeed,  the helicity mixing angle, $\theta_B$, which determines
the mixing between the left and the right
components (in ultrarelativistic case) is
\begin{equation}
\tan 2\theta_B = \frac{2 \mu B}{V_{SF} - \frac{\Delta m^2}{2E} } .
\end{equation}
(In medium with constant parameters $\theta_B$ determines the depth of
precession,  when the initial state has a definite chirality:
$A_P = \sin^2 2\theta_B$).  Since in general the field strength depends
on distance,  the angle  $\theta_{B}$ can be large  enough  in a
wide spatial region provided
 $\mu B \sim V_{SF} \gg \Delta m^2/2E$. For large  $B$
the spin-flip effect may not be localized
in the resonance layer, and moreover, the resonance itself may not be local
\cite{APS}.

The rapidity of the $\theta_B$ change
on the way of neutrino, $\dot{\theta}_B \equiv {\rm d}\theta_B /{\rm d}r$,
determines
the adiabaticity condition for spin-flip:
$$
\pi \dot{\theta}_B {\buildrel < \over {_{\sim}}}
\Delta H = \sqrt {(2\mu
B)^2 + \left(V_{SF} -
\frac{\Delta m^2}{2E}\right)^2}.
$$
The adiabaticity parameter can be
written as
\begin{equation}
\kappa \equiv \frac{\Delta H}{\pi \dot{\theta}_B} =
\frac{(V^2 + (2\mu B)^2)^{3/2}
}{
\pi (\mu \dot{B} V - \dot{V} \mu B)} ,
\end{equation}
here $V \equiv V_{SF} - \Delta m^2/2E$.
In resonance, $V = 0$ or $V_{SF} \cong \Delta m^2/2E$, the expression (18)
reduces to (10).
Beyond the resonance $\kappa$ depends on the magnetic field
change. In the extreme cases one has
\begin{equation}
\kappa \approx \left\{
\begin{array}{l}
\kappa_R
\left(\frac{V_{SF}}{\mu B}\right)^3, ~~~ V_{SF} \gg \Delta m^2/2E\\
\kappa_R \left(\frac{\Delta m^2/E}{\mu B}\right)^2
\left(\frac{\dot{V}_{SF}}{\mu \dot{B}}\right),~~~ V_{SF} \ll
\Delta m^2/2E
\end{array}
\right. ,
\end{equation}
where $\kappa_R$ is the resonance adiabatic parameter
for a given point (10).  According to (19) the adiabaticity condition
is strongly relaxed, if
$V_{SF} \gg \mu B$ above the resonance,  and if
$\Delta m^2 /2E \gg \mu B$ below the resonance. For the
density and the field profiles under consideration  these inequalities
are fulfilled,  so that the adiabaticity is the most crucial
in the resonance.

If the initial and final mixings are small:
$\theta_B^i \approx \pi/2$ and $\theta_B^f \approx 0$
and the level crossing is  local, one can estimate the
survival probability using the Landau-Zener formula:
\begin{equation}
P \approx P_{LZ} = e^{-\frac{\pi^2}{4} \kappa_R} =
e^{- \frac{\pi^2}{4}\left(\frac{B}{B_A}\right)^2},
\end{equation}
where $B_A^2 \equiv \pi \dot{V}/8 \mu^2$ (see (10)).
At the adiabatic condition, $B = B_A$, one gets $P_{LZ} = 0.085$.
The probability increases quickly with decrease of the field:
$P_LZ = 0.3,\;  0.54,\; 0.8$,  when  $B/B_A = 0.7,\; 0.5,\; 0.3$
correspondingly (see Fig.~3). An appreciable effect
could be  for $P_{LZ} = 0.8$, when the magnetic field is
about 3 times smaller than the adiabatic value.

\subsection{The precession and the
adiabaticity bounds}

The propagation of neutrinos with magnetic moment
in the magnetic field and matter is an  interplay of two processes:
the precession and the resonance spin conversion.
A relation
between them is determined largely by density distribution. Let us
define the {\it precession bound} for the product $(\mu B)_{P}$ as
\begin{equation}
 \mu B_P = \frac{1}{2} V_{SF}.
\end{equation}
At $(\mu B) = (\mu B)_{P}$, the precession
may  have the depth $A_{P} = 1/2$ (see (6)); for $(\mu B) \ll  (\mu
B)_P$ the precession effects can be neglected.
Let us also introduce the
{\it adiabaticity bound} $(\mu B)_{A}$
using the adiabaticity condition (10):
\begin{equation}
 (\mu B)_{A} = \sqrt{\frac{\pi \dot{V}_{SF}}{8}} \approx
\sqrt{\frac{\pi}{8 H_{SF}}}
V_{SF}^{1/2},
\end{equation}
where $H_{SF} \equiv [{\rm d}V_{SF}/{\rm d}x]^{-1} V_{SF}$
is the typical scale of
the potential change. For $(\mu B) >
(\mu B)_{A}$ the  level crossing is adiabatic.
Comparing (21) and (22) one finds that for fixed
$H_{SF}$ the precession bound  increases with $V_{SF}$
faster than the adiabaticity bound. Consequently, for large $V_{SF}$,
the inequality  $(\mu B)_{P} \gg  (\mu B)_{A}$
holds and the conversion
dominates over precession. The suppression of  $V_{SF}$
makes the precession effect more profound.
If the potential changes as $V_{SF} \propto r^{-3}$, one gets
\begin{equation}
 (\mu B)_{A}\, \propto \, \frac{1}{r^{2}}, ~~~~~~~~ (\mu B)_{P}\,
 \propto \frac{1}{r^{3}}.
\end{equation}
Clearly, $(\mu B)_{P} < (\mu B)_{A}$ in the
hydrogen envelope, and $(\mu B)_{P} > (\mu B)_{A}$ in the center
of star. The adiabatic conversion dominates in the inner parts,  whereas
the precession may dominate in the external layers of star
(Fig.~3). In the isotopically neutral region
one has $(\mu B)_P \sim  (\mu B)_{A}$,
i.e., both processes may be  essential (see Sect.~4). In fact,
the condition (21) is not sufficient for a
strong spin-precession effect. The spatial size of the region where (21) is
fulfilled, $\Delta r$,  should be large enough, so that the inequality (1)
or  $\Delta r {\buildrel > \over {_{\sim}}}
{1}/{\mu B}$ is satisfied.
For $r > 0.3 R_{\odot}$ the latter  bound is even stronger than (21).
Instead of $\mu B$, in further discussion we will give  the
estimations of  the magnetic fields
at $\mu = 10^{-12}\mu_B$.
Rescaling to other values of $\mu$ is obvious.

The adiabaticity and the precession  bounds  determined by the density
distribution should be compared with the magnetic field profile
of  star. In fig. $3$, we depict the  profile (2)
with $k = 2$ and $B_{0} = 1.5\cdot 10^{13}$ Gauss.
The strength of this field is practically everywhere
below both the adiabatic, $B_A$, and the precession, $B_P$, bounds
($\mu = 10^{-12}\mu_B$) which means
that  spin-flip effects are rather weak. The effects of the global
field (2) may be strong,  when  $B_0 \geq  10^{14}$ Gauss.
However, for $k = 3$ even extremely large central field, $B_{0} \sim
10^{17}$ Gauss,
corresponds to very weak magnetic field in the isotopically neutral
region. Local mechanisms of
the magnetic fields generation  due to possible convection and
differential rotation may produce the local field being much
larger than the global field, as in the case of the Sun.

\subsection{Effects of  jumps of the potential}

The electron density number $n (2Y_{e}-1)$, and consequently, the effective
potential $V_{SF}$ change quickly in the ignition layers,
so that for reasonable values of $\mu B$ the adiabaticity is broken.
 This feature is
reflected by peaks in the  adiabaticity bound (fig.~3).
To a good approximation one can consider these changes as
sudden jumps of the potential, $\Delta V_j \equiv
V_{\rm max} - V_{\rm min}$,  at fixed points $r_j$.
The effect of jump reduces to  jump-like change
of the helicity mixing angle $\theta_B$.

The  jumps  distort the energy dependence of probabilities.
The position of the energy interval  affected by the jump is related via the
resonance condition to the values of
$V_{\rm min}$ and $V_{\rm max}$.  The character of distortion
depends on the magnitude of jump,  $\Delta V_{SF}$, as compared with
the energy width of the resonance layer $\Delta V_{R}$
($\Delta V_{R} = 2\mu B$).
(See a similar consideration for the flavor case in \cite{KS}). If $\Delta
V_{j}  \leq \, \Delta V_{R}$, the jump leads to the
peak of the survival probability with width, $\Delta E$,  determined
by the width of the resonance layer: $\Delta E/E \sim  2\mu B/V_{SF}(r_j)$
\cite{KS}.
The height of the peak is given by the ratio:
$$
\Delta P \sim
\left(\frac{\Delta V_j}{\Delta V_{R}}\right)^{2}
= \left(\frac{\Delta V_j}{2\mu B}\right)^{2}.
$$
For  $2 \mu B \gg \Delta V_j$ the jump effect is smoothed:
since the size of the resonance region is large,
the  width of peak is also large,
but  its  height  turns out to be  suppressed.

If $\Delta V_j  \geq  2\Delta V_{R} \approx \mu B$, the height of
the peak is of the order  $1$, and the  width
is fixed by the size of jump:
\[
\frac{\Delta E}{E} \sim  \frac{\Delta V_j}{V_{SF}}~~.
\]
For $\Delta V_{SF} \gg  \Delta  V_{R}$, the jump
suppresses the transition in the interval
with minimal and maximal energies
determined by $V_{\rm max}$ and $V_{\rm min}$.

The most profound effect appears,   when
without jump the adiabaticity is fulfilled and the transition is practically
complete. If the width of the resonance is small
($2\mu B < V_{SF}$) and if $\Delta V_j \sim \Delta V_R$, the jump
induces  a thin peak of the survival probability. Such conditions
can be fulfilled in the inner part of the isotopically neutral region.

Note that in the considered model
of the star there are two big jumps of the potential
($\Delta V_j/V_{SF} \sim  10$)
at $r
=  7\cdot 10^{-2}R_{\odot}$ and $r =  0.3 R_{\odot}$ (fig.1).
Also small jumps ($\Delta V_j/V_{SF}
{\buildrel < \over {_{\sim}}}     1$) exist.

The elements of the dynamics presented
here allow to understand
all features of the evolution of the neutrino state.

\section{Neutrino propagation in isotopically neutral region}

Let us consider for definiteness the spin-flavor transitions
$\bar{\nu}_e - \nu_{\mu}$ and $\nu_e - \bar{\nu}_{\mu}$.
If   $\Delta m^{2} \leq 0.1~{\mbox{eV}^{2}}$ the
level crossing takes place in the isotopically neutral region.
For realistic magnetic fields (profile (2) with $B_0 \ll 10^{17}$
 Gauss) the effect of  $\bar{\nu}_e - \nu_{\mu}$ mixing
can be neglected in the inner part of the star. Therefore,
the spin-flavor evolution of the neutrino state starts from
the inner part of the isotopically neutral region $r_i \sim
10^{-3}R_{\odot}$.
 At $r_i$
the states  coincide practically with pure helicity/flavor states.
Flavor conversion, if efficient,  can be taken into account as
further independent transformation (sect.~5).
Also the transitions which involve  $\nu_{\tau}$  can be considered
separately.

\subsection{Spin-flip effects in global magnetic field}

We use the  profile (2) with $k =2$ and  take different values
of $B_0$.
For $B_{0} {\buildrel < \over {_{\sim}}}
  2\cdot 10^{13}$ Gauss
the magnetic field strength is below the
precession bound ($\mu = 10^{-12}\mu_B$) in the isotopically neutral
region.  Consequently, the depth of the precession is
suppressed, and  the transition has a local resonance character (fig.~4).
The transition occurs  only in the resonance region
whose  width is determined by  $\Delta V_{R}/V_{SF} = 2\mu B/V_{SF}$.
The efficiency of the transition  depends on $\kappa_R$.
The strongest effect takes place  for the energies (over $\Delta m^2$),
$E/ \Delta m^2$,  which correspond to the weakest violation
of the resonance adiabaticity
($B(r)$ is closer to the adiabaticity bound (fig.~3, 4)).
With the increase of $B_0$ the  $E/ \Delta m^2$ range of
strong conversion expands (fig.~5).

If $B_0 {\buildrel > \over {_{\sim}}}
 5 \cdot 10^{13}$ Gauss, the field profile is above both the
precession and the adiabatic bounds, and the  conversion becomes essentially
nonlocal (fig.~4). For $E/\Delta m^{2} >  10^{3}$ MeV/eV$^2$
initial mixing is still very small: $\theta_B \approx \pi/2$,
$\sin^2 2\theta_B \approx 0$. Therefore, the neutrino state
coincides with one of the eigenstates in medium.
Since the adiabaticity is fulfilled, the survival probability is uniquely
determined by the helicity mixing angle: $P(\nu_e \to \nu_e) \approx \cos^2
\theta_B$.
The precession effects are very  small (fig.~4).
Before the resonance the angle is determined  from
$\tan 2 \theta_B \approx \mu B(r)/ V_{SF}(r)$.
The mixing increases, when the neutrino enters the
isotopically neutral region,
and if $\mu B \gg V_{SF}$ the probability approaches $P \sim 1/2$ (fig.~4).
 After the resonance layer the angle diminishes quickly:
$\tan2 \theta_B \sim 2 \mu B /(\Delta m^2/2E)$,
since the magnetic field decreases with distance,
whereas the splitting approaches the asymptotic value $\Delta m^2/E$.
Correspondingly, the probability goes to zero and one has almost
complete spin-flip with  $P(\nu_e \to \nu_e) \approx 0$ (fig.~4).

For large $E/ \Delta m^2$
the resonance is in  the external
layers of the star, where the adiabaticity is broken. The final survival
probability increases with $E/ \Delta m^2$,  approaching
the  value of probability
before the resonance.  As the result one gets
$P {\buildrel > \over {_{\sim}}}    1/2$.
At very large
$E/ \Delta m^2$ mass splitting becomes unessential and the dependence
of probability on $E/ \Delta m^2$ disappears.

The bump in the survival
probability $P(E)$ in the energy interval
$E/\Delta m^2 = (10^7 - 10^8)$ MeV/eV$^2$
(fig.~5) is due to a density jump at $r = (0.1 - 0.5) R_{\odot}$
(fig.~4 a). For very large $\mu B$ the effect of jump
is suppressed according to the discussion in sect. 3.4.

\subsection{Spin-flip in local magnetic field}

Let us suppose that a strong magnetic field exists
in some spatial region $\Delta r$  between $r_1$ and $r_2$,
so that the transitions take place in $\Delta r$ only.
For simplicity we  consider a constant magnetic field
$B_c$.
If $\mu B$ is below the precession bound, the conversion occurs
in the resonance layer, and the transition probability is determined by
$\kappa_R$. The effect is  appreciable
in the energy interval determined via the resonance condition by  the
maximal and minimal values of
the potential in the layer with magnetic field (fig.~6).
When the adiabaticity is broken,  the
maximal effect corresponds to the resonance in the central part of
the region with the magnetic field.
In this case there is no averaging over the
precession phase and  smooth dependence of the probability
on energy can be modulated.
The energy dependence can be distorted also by
jumps of the potential.

If the field is strong enough and  the adiabaticity condition is
fulfilled, the average probability follows the change of helicity mixing
angle $\theta_B(r)$ (17).
The survival probability,  as a function of
$E/ \Delta m^2$ is determined by the adiabatic formula:
\begin{equation}
P_{\nu_e \to \nu_e} = \frac{1}{2}\left(1 + \cos 2 \theta_B^i \cos 2
\theta_B^f\right)
+ \frac{1}{2} \sin 2\theta_B^i \sin 2\theta_B^f \cos \Phi ,
\end{equation}
where $\Phi$ is the   precession phase:
\begin{equation}
\Phi = \int_{r_1}^{r_2} {\rm d}r \sqrt {(2\mu B)^2 + (V_{SF} -
\Delta m^2/2E)^2}.
\end{equation}
Here $\theta_B^i$ and $\theta_B^f$  are the values of the mixing angle
(17) at  $r_1$ and  $r_2$.
For $\mu B \gg V_{SF}$, the influence of matter is small and the
survival probability is large for $E/\Delta m^2 > 1/(\mu B)$.
If $E/\Delta m^2 \gg 1/(\mu B)$ the precession has maximal depth
(fig. $6$).

\subsection{On the energy dependence of probability}

The survival probability as function of the neutrino energy
has some common features for different magnetic field profiles.
There are   two characteristic energies:
$(E/\Delta m^2)_{\rm min}$ and
$(E/\Delta m^2)_{\rm max}$.
The former,  $(E/\Delta m^2)_{\rm min}$, is determined by the adiabaticity
violation in the inner part of the isotopically neutral region, or in the
inner part of the region with magnetic field (in the case of the local field).
The latter, $(E/\Delta m^2)_{\rm max}$, is fixed in such a way that for
$E > E_{\rm max}$ the effect of mass splitting can be neglected.
For the local field $(E/\Delta m^2)_{\rm max}$
is determined by the
potential at the outer edge of the region with magnetic field.
In the case of global field one has
$(E/\Delta m^2)_{\rm max} \approx 10^{-10}$ eV$^2$
which follows from the adiabaticity violation
due to a fast decrease of $\mu B$
or $V_{SF}$ at the border of the isotopically neutral region
(fig.~3).

The energies ($E/\Delta m^2)_{\rm max}$ and $(E/\Delta m^2)_{\rm min}$
divide the whole energy interval into three parts.

1).  Region of matter/vacuum suppression of the spin-flip:
$E/\Delta m^{2} < (E/\Delta m^{2})_{\rm min}$.
For neutrinos with such energies before the resonance layer the precession
amplitude  is strongly suppressed by
matter effect. There is no appreciable conversion in the resonance region,
since the adiabaticity is strongly broken; after the resonance the
amplitude is suppressed by vacuum splitting: $A_{P}\, \sim \, (\mu
B)/(\Delta m^{2}/E)$. As a result, the
probability of spin-flip is small.

2).  Resonance region:  $(E/\Delta m^{2})_{\rm max} <
E/\Delta m^{2} < (E/\Delta m^{2})_{\rm min}$.
The resonance is in the isotopically
neutral region. If  the adiabaticity is unbroken  or
weakly broken, the survival probability is small.
Smooth energy dependence
of $P$ due to conversion can  be
 modulated by the effect of density jumps, as well as
by the precession effect,
if e.g. the initial mixing angle is not small.

3).  Precession (asymptotic) region:
$E/\Delta m^{2} > (E/\Delta m^{2})_{\rm max}$.
Here the  mass splitting ($\Delta m^2/E$) can be neglected,
and therefore the spin-flip probability does not depend on energy.
The spin-flip is due to precession
and possible adiabatic (non-resonance) conversion.

If the layers  with magnetic field are in the inner part of the
isotopically neutral region, the
resonance effects can be more profound. Here the adiabaticity condition
is fulfilled, even when the width of the resonance is
small, $2\mu B/V_{SF} < 1$. For the external layers with small density the
adiabaticity holds for $2\mu B/V_{SF} > 1$ only, i.e. when
the width of resonance is large.

\section{Spin-flip and other transitions }

\subsection{Spin-flip and flavor conversion}

Let us consider the effects  of the flavor conversion
$\nu_e - \nu_{\mu}$ in addition to the spin-flip. In the isotopically
neutral region the positions of the spin-flavor resonance, $r_s$,
and the flavor resonance, $r_f$, are strongly
separated in space: $r_{s} \ll r_f$.
If the flavor mixing as well as the magnetic field strength are
sufficiently small, both flavor  and spin-flavor transitions are
local.
The crossings of flavor and spin-flavor
resonance layers are independent, and the total transition probability
is the product  of the flavor and the spin-flavor probabilities.

Three neutrino states, $(\nu_e, \bar{\nu}_e, \nu_{\mu})$,
are involved in the transitions.
Correspondingly, one can introduce
3$\times$3 matrix of probabilities $S$ which
relates the original,
$F^0 = \left(F^0(\nu_e), F^0(\bar{\nu}_e), F^0 (\nu_{\mu})\right)$,
and the final, $F = \left(F(\nu_e), F(\bar{\nu}_e), F(\nu_{\mu})\right)$,
fluxes:
\begin{equation}
F = S F^0.
\end{equation}
Let $r'$ be some point between the two
resonances: $r_{s} \ll r' \ll r_f$.
Then  in the region $r < r'$
the matrix $S$ depends only on the spin-flavor transition
probability, $P_s(\bar{\nu}_e - \nu_{\mu})$:
$S = S_s(P_s)$. (Evidently,  $(1 - P_s)$
coincides with the survival probability calculated in the sect.~4).
In the region $r > r'$ the matrix $S$ depends on the flavor
transition probability, $P_f
(\nu_e - \nu_{\mu})$ transition: $S = S_f(P_f)$.
The total matrix is  the product $S = S_s(P_s) \times S_f(P_f)$:
\begin{equation}
S =
\left(
\begin{array}{lll}
(1 - P_f) & P_f P_s        & P_f(1- P_s) \\
0         & (1 - P_s)      & P_s  \\
P_f       & (1 - P_f) P_s  & (1 - P_f)(1- P_s)
\end{array}
\right) .
\end{equation}

Let us consider the dependence of $S$ on energy.
If the flavor mixing is small (say $\sin^2 2\theta < 0.01)$,
and  the  magnetic field is situated
in the external part of the star, the energy regions of the strong flavor
transition and the strong spin-flip effect do not overlap.
The neutrinos of high energies  flip the  helicity,
whereas low energy neutrinos undergo the flavor transition.
The overlap of the resonance  regions for the spin-flip and the
flavor conversion takes place for sufficiently large mixing angles
or/and  if there is a strong enough magnetic field in the inner part of
star:  $r < 0.3R_{\odot}$ (Fig.~$7$).
In the overlapping energy region the probabilities $P_s$ and  $P_f$
differ from zero and the  electron neutrinos undergo a double transition:
$\bar{\nu}_e \rightarrow \nu_{\mu} \rightarrow \nu_e$ (fig.~7).

\subsection{Effects of tau neutrino}

Inclusion of  $\nu_{\tau}$   adds
two new  level crossings: the
flavor, ($\nu_e - \nu_{\tau}$), and spin-flavor, ($\bar{\nu}_e -
\nu_{\tau}$) ones (fig.~2). Other crossings involving
$\nu_{\mu}$ and $\nu_{\tau}$ do not give observable effects.
Due to the assumed mass hierarchy the transitions
with  $\nu_{\tau}$ (the inner region of the star)
can be considered independently from the transitions involving
$\nu_{\mu}$  (outer part). Consequently, the  total
transition matrix is the
product of the transition  matrices in the inner and the outer regions.

Three states, $\nu_e, \bar{\nu}_e, \nu_{\tau}$,
are transformed  in the inner part (fig. 2).
For monotonously decreasing  density, the order of
resonances is the same as before: first --
spin-flavor and then flavor one. If the $\nu_{\tau}$ resonances
are enough separated, the transition matrix can be found from
that in (27) by substitution
$P_f \rightarrow P'_f$ and $P_s \rightarrow P'_s$, where
$P'_f$ and $P'_s$ are the $2\nu$-probabilities of flavor
and spin-flavor transitions in the inner region.
(In fact one should consider $4 \times 4$ transition matrices
in the ($\nu_e$, $\bar{\nu}_e$, $\nu_{\mu}$, $\nu_{\tau}$)-basis in the
inner and the outer parts and find their product.)
The effects due to the  $\nu_{\tau}$-resonances can be
taken into account as the initial conditions,$F^i$, for the fluxes
at the border with the outer region.
Using the  matrix (27) with substitution mentioned above
we get
\begin{equation}
\begin{array}{l}
F(\nu_{e})^i =
(1 - P'_f)F^0(\nu_e)  + P'_f P'_s
F^0 (\bar{\nu}_e) + P'_f(1- P'_s) F^0(\nu_{\tau}),\\
F(\bar{\nu}_{e})^i =  (1 - P'_s) F^0 (\bar{\nu}_e) +
P'_s F^0(\nu_{\tau}), \\
F(\nu_{\mu})^i = F^0 (\nu_{\mu}).
\end{array}
\end{equation}

Further on we will concentrate on the cosmologically interesting
values $m_3 \sim (2 - 7)$ eV. This corresponds
to $\Delta m^2 = (4 - 50)$ eV$^2$, so that for $E \sim 20$ MeV the
resonances are in the inner part of the isotopically neutral region:
$r \sim (1 - 2)\cdot 10^{-3} R_{\odot}$. Here the
total density is $(0.3  - 3)\cdot 10^8$ g/cm$^3$ and $(1 - 2 Y_e)
\sim (2 - 4)\cdot 10^{-2}$. According to fig.~3,
the precession bound equals
$(10^{13} - 3\cdot 10^{15})$ Gauss, and the adiabaticity
bound is $(0.5 - 2) \cdot 10^{11}$ Gauss. Thus an appreciable spin-flip
transition implies very large magnetic field:
$B > 10^{10}$ Gauss. In contrast, the flavor transitions
can be efficient even for very small mixing angles:
$\sin^2 2\theta > 10^{-6}$.
For  $\Delta m^2$ under consideration the spin-flip effects may be more
probable below the isotopically neutral region.

\subsection{Transitions in the central region of the star}

The  collapse and the shock wave propagation lead to strong
time dependence of the density profile in the central region of the star.
Moreover, the profile may be non-monotonous.
With approaching the center of star the total matter density may
first diminish from $10^8$ g/cm$^3$
to $(10^6 - 10^7)$ g/cm$^3$ at $r \sim 100$ km, and then
rapidly increase at the surface of the
protoneutron star,  $r < (20 - 30)$ km. Moreover, in the central part
the neutrino-neutrino scattering should be taken into
account \cite{pant}.
Due to the non-monotonous behavior of $V_{SF}$,  for $\Delta m^2 = (4 -
50)$ eV$^2$ the additional level crossings  appear in the
region  $r \sim (30 - 100)$ km:
second flavor crossings   $\nu_e - \nu_{\tau}$,
and second  spin-flavor crossings $\bar{\nu}_e - \nu_{\tau}$.
The order of resonances from the center is the following:
spin-flavor (s), flavor (f), flavor (f), spin-flavor (s) ---
$s, f, f, s$.
Note that now there is some part of the profile where the potential
increases with distance, and consequently, the order of resonances is
reversed: neutrino crosses first the flavor resonance and
then the spin-flavor one.
This may have important consequences for observations (see sect.
6.4).

At distances (30 - 100) km the adiabaticity bound for the
spin-flip equals $B_A = (1 - 2) \cdot 10^{11}$ Gauss.
This is only slightly higher than that in the inner edge of
the isotopically neutral region.
However, the existence of such a field
is more probable in the inner part; for the profile (2) with $k
= 2$ and $B_0 = 10^{13}$ Gauss the field is only 2 - 3 times
smaller than the adiabaticity bound.
Thus the magnetic moment  $\mu \sim 10^{-12}\mu_B$ may give strong
spin-flip effect here.

If in the inner part all four transitions are efficient, there is no
observable effect: the final state will coincide with the initial one.

\section{Implications}

In principle,  future experiments will give information on the energy
spectra of
different neutrino species. Confronting these spectra with each other, one
can get essentially model independent information about possible neutrino
transitions.
It is convenient to introduce  three types of original spectra:
soft, $F_s(E)$, middle, $F_m(E)$, and hard,
$F_h(E)$, which coincide with original $\nu_e$-,
$\bar{\nu}_e$-,
and $\nu_{\mu}$-spectra, respectively. The  muon and tau neutrinos
 and antineutrinos
are  indistinguishable in the standard electroweak model
at low energies, and will be detected by neutral currents. Therefore
we will consider the total flux of the ``non
electron neutrinos",
$F(\nu_{ne})$. If there is no neutrino transitions,
then at the exit:
\begin{equation}
F(\nu_e) = F_s,~~~ F(\bar{\nu}_e) = F_m,~~~
F(\nu_{ne}) = 4 F_h.
\end{equation}
Let us find the signatures of the spin-flip transitions,
as well as the sensitivity of the  $\nu$-burst studies to the neutrino
magnetic moments.
We estimate the values of $B$ (at $\mu = 10^{-12}\mu_B$) needed for
different effects to be observable.

\subsection{Sensitivity to magnetic moments and magnetic fields}

The suppression of the effective potential by $3-4$ orders of magnitude
diminishes the strength of the magnetic field (magnetic moment)
needed for a strong spin-flip effect. The adiabatic and the precession
bounds decrease by $1.5-2$ and $3-4$ orders of magnitude respectively.
We take as the criterion of the sensitivity of the
$\nu$-burst studies to the spin-flip effects the magnitude of the
magnetic
field strength $B_{s}$ ($\mu = 10^{-12} \mu_B$) at which
the probability of transition is $\sim 1/2$
at least for some neutrino energy interval.
According to Fig.~$2$, in the most part of the isotopically neutral region
($r \leq 0.3 R_{\odot}$) the adiabaticity  bound is below
the precession bound. Only in the external part of star the
precession could be more preferable. Consequently, for
$r \leq 0.3 R_{\odot}$ the sensitivity limit corresponds to
inequality $2 \mu B < V_{SF}$. The latter means that
mixing is small everywhere apart from the resonance
layer and  the transition
is due to the level crossing with not so strong adiabaticity violation.
The nonaveraged over the precession phase probability $P \sim 1/2$, can
correspond to the value of the Landau-Zener probability $P_{LZ} \sim 0.75$.
Therefore we will define
the sensitivity limit $B_s(r)$ in a given point as the
strength of the magnetic field,
for  which   $P_{LZ}(r) \sim 0.75$. According to the estimations in section
3.2,
the sensitivity bound can be
about 3 times  smaller than  the adiabaticity bound:
$B_s \approx B_A/3$ (fig.~3).

For the effect to be appreciable,  the size of the region with
magnetic field, $\Delta r_B$, should be comparable with the size of
the resonance region. If $V_{SF} \propto r^{-3}$,
we get  $\Delta r_B = ({\rm d} V_{SF}/{\rm d}r)^{-1}2\Delta V_R
\approx 4\mu B r/3V_{SF}$,
and in particular, for $2 \mu B /V_{SF} < 0.3$:  $\Delta r_B < 0.2 r$.
In fact, $B_s$ should be considered as the average field
in the region $\Delta r_B$.

The transition takes place in the resonance region whose  position
depends via the resonance condition on
$E/\Delta m^2$. Therefore for fixed
$E/\Delta m^2$ one can define the sensitivity limit $B_s$ in
a certain region of star $r = r(V_{SF}(E/\Delta m^2))$.
For $\Delta m^2 = (0.3 - 1) \cdot 10^{-5}$ eV$^2$ ($E \sim 20$ MeV)
which corresponds to the MSW solution to the solar neutrino problem
\cite{MS,kras} the spin-flip resonance lies at
$r \sim (1 - 3) \cdot 10^{-2}R_{\odot}$,
and according to fig.~2, the sensitivity limit equals $B_s = (2 -
5)\cdot10^{6}$ Gauss. For the oscillation solution of the
atmospheric neutrino problem ($\Delta m^2 \sim  10^{-2}$ eV$^2$)
\cite{atm} the resonance is at $r \sim (2 - 3)\cdot10^{-3}~R_{\odot}$,
and $B_s \sim 10^9$ Gauss.

In the external part  of the isotopically neutral region,
$r = (0.3 - 0.7)~ R_{\odot}$,
the precession bound is essentially below the
adiabaticity bound. However, here for $2\mu B \sim V_{SF}$
the precession length
is already comparable with the distance
to the center of star and the condition
(3) becomes more important. At $r \approx  0.3 R_{\odot}$ one gets
$B_s \sim (2 - 3) \cdot 10^4$ Gauss.  This region corresponds to values
 $\Delta m^2 = (10^{-9} -  10^{-8}$) eV$^2$ which
are interesting from the point of view of the resonant spin-flip in the
convection zone of the Sun.

In the range $r \approx  (0.7 - 1) R_{\odot}$ the matter effect can be
neglected and the sensitivity is determined by vacuum precession:
$B_s > 2 / R_{\odot} \sim (2 - 3)\cdot 10^{4}$ Gauss.
This number can be compared with  value of $B$ needed to solve the
solar neutrino problem.
For strong $\nu_{eL}\rightarrow \bar{\nu}_{\mu R}$
conversion in convection zone of the sun $(r \sim \, (0.7-1)R_{\odot})$
one needs the magnetic field as large as $B \approx  3\cdot 10^{5}$
Gauss \cite{mfs}. In supernova, at
the same distance from the center, the field of about $10^{4}$
Gauss is enough. Conversely, if the magnetic field is $\sim 3\cdot
10^{5}$ Gauss, the $\nu$-burst will be sensitive to $\mu \, \sim \,
3\cdot 10^{-14}  \mu_{B}$.

Note that in the isotopically neutral region
one needs for appreciable $\nu_{eL}-\bar{\nu}_{\mu R}$ conversion
$1.5-2$ orders
of magnitude smaller field than  for conversion into a sterile state:
$\nu_{eL}\rightarrow \nu_{sR}$.

Comparing the above  results with those of sect.~5.3 we find
that the sensitivity to $\mu$ can be  even higher
in the isotopically neutral region than
possible sensitivity to $\mu$
in the central region of  star. (Of course, the latter corresponds to large
values of
$\Delta m^2$).

\subsection{Bounds on $ \mu B$ from SN1987A}

The spin-flip $\nu_{\mu}\rightarrow \bar{\nu}_{e}$
leads to an appearance of the high energy tail
in the $\bar{\nu}_{e}$-spectrum at the Earth:
\begin{equation}
F(\bar{\nu}_e) = (1-P_s) F_m + P_s F_h~,
\end{equation}
where $P_s$ is the transition probability.
At small mixing angles the flavor conversion does not change
this result (see fig. $3$). Note that in (30)
$P_s$ can be close to 1, in contrast with
averaged vacuum oscillation effect which gives $P \leq 1/2$.

The absence of the distortion of the $\bar{\nu}_e$-energy
spectrum, and
in particular,  the absence of the high energy tail,
gives the bound on $\mu B$ as function of $\Delta m^2$.
For a class of supernovae models the  data from SN1987A allow to get
the restriction   $P_s < 0.35$
under the assumption that $P_S$ does not depend on energy
\cite{SSB}. In fact,
at the border of sensitivity the suppression pit is rather thin
(figs.~5, 6) and the energy dependence can not be
completely neglected. In this case the  strongest
observable effect takes place, when a position of the
 pit coincides with the high energy part of
the $\nu_{\mu}$-spectrum. The sensitivity limit obtained in sect.~6.1
gives an
estimation of the upper bound on $\mu B(r)$. The value of the potential
$V_{SF}(r)$ in the corresponding point $r$
determines via the resonance condition the value of $\Delta m^2$.
For example, at $\Delta m^{2} \simeq 10^{-8}$ eV$^{2}$ (resonance at
$r  =  0.3R_{\odot}$) one gets the upper bound as
$B (0.3 R_{\odot}) <  2\cdot 10^{4}$ Gauss.
Similarly, if $\Delta m^{2}  =
10^{-6}$ eV$^{2}$: $B(0.1 R_{\odot}) < 3\cdot 10^{5}$ Gauss,
and if $\Delta m^{2} \sim  3\cdot 10^{-5}$ eV$^{2}$:
$B(0.01 R_{\odot}) < 10^{7}$ Gauss etc.. In the case of the global
magnetic field (2) with $k =2$ these bounds correspond to the bound on
the field at the surface of the protoneutron star
$B_0 {\buildrel < \over {_{\sim}}}    10^{13}$ Gauss
($\mu = 10^{-12}\mu_B$).

\subsection{Transitions of the degenerate or massless neutrinos}

The  mass differences
$\Delta m^{2} {\buildrel < \over {_{\sim}}}
10^{-10}$ eV$^{2}$ can be neglected for
the neutrino energies $E \sim 5 - 50$ MeV.
This corresponds to the asymptotic  region
in $P$-dependence on $E/ \Delta m^2$ (fig.~4, 5).
As we have noticed in sect.~2, for $\Delta m^2 \approx 0$  there are two level
crossings $\nu_e - \bar{\nu}_{\mu}$ and $\bar{\nu}_e - \nu_{\mu}$  at the
same point, $r \sim  R_{\odot}$. The crossings are  induced by
change of the nuclear  composition
on the border between the H-envelope and the ${}^{4}$He-layer. However,
in this region  the
effective density changes  very quickly, and  the adiabaticity
condition implies very strong magnetic
field $B((0.7-1) R_{\odot}) > 3\cdot
10^{5}$ Gauss which exceeds the precession bound.
The effect is mainly due to the precession, and
the field should be as strong as $B \sim (2 - 5)\cdot10^4$ Gauss.
The probabilities of transitions $\nu_{e}\leftrightarrow \bar{\nu}_{\mu}$
and $\bar{\nu}_{e}\leftrightarrow \nu_{\mu}$ are  equal
and do not  depend on energy. The final
spectra of $\nu_e$ and $\bar{\nu}_e$ equal:
$F(\nu_{e}) = (1 - P_s) F_s + P_s F_h$,
$F(\bar{\nu}_{e}) = (1 - P_s) F_m + P_s F_h$.
In the case of complete
transformation, $P_s \approx 1$, the
$\bar{\nu}_{e}$- and $\nu_{e}$- energy spectra coincide with the
initial $\nu_{\mu}(\bar{\nu}_{\mu})$-spectrum:
\begin{equation}
F(\nu_{e}) = F(\bar{\nu}_{e}) = F_h ~~~
\end{equation}

For $\Delta m^{2} >  (10^{-9}-10^{-8})$ eV$^{2}$
(which is interesting for the spin-flip of solar
neutrinos \cite{Kh})
the effect of the mass splitting becomes important. The
probabilities of $\bar{\nu}_{e}\rightarrow \nu_{\mu}$
and $\nu_{e}\rightarrow \bar{\nu}_{\mu}$ transitions
are different, and depend on energy.

\subsection{Distortion of neutrino energy spectra}

Using the matrix of the transition probabilities
for the isotopically neutral region (27) and the
initial conditions (28) we get for the final neutrino spectra:
\begin{eqnarray}
F(\bar{\nu}_e) &=&  (1 - P_s)(1 - P'_s) F_m +
[(1 - P_s)P'_s +   P_s] F_h~ ,\nonumber\\
F(\nu_e)     &  =&  (1 - P_f)(1 - P'_f)F_s +
[P'_f P'_s (1 - P_f) +  P_f P_s (1 - P'_s)] F_m
\nonumber \\ && \mbox{}
+ [(1 - P_f) P'_f (1 - P'_s) + P_f(P_s P'_s + 1- P_s)] F_h~,\\
F(\nu_{ne}) &=& a_s F_s + a_m F_m + a_h F_h~,
\nonumber
\end{eqnarray}
where
\begin{eqnarray}
a_s &=& P'_f +  P_f (1- P'_f)\nonumber \\
a_m  &=& P_f P'_f P'_s +  (1 - P'_f) P'_s  +  (1 - P_f) P_s (1- P'_s)\\
a_h &=& 4 - a_s - a_m \nonumber .
\end{eqnarray}
The important conclusions can be drawn  from
(32, 33) immediately.  The $\bar{\nu}_e$-spectrum is a mixture of
the middle (the original $\bar{\nu}_e$ - spectrum)
and hard components. It does not acquire soft
component,  except for the case of the
non-monotonous change
of the density in the central part of the star (see latter).
The spectrum depends on the spin-flip probabilities $P_s$ and $P'_s$
and does not depend on  probabilities of the flavor transitions.
The changes are stipulated by the
spin-flip effects only. Thus the distortion of
the $\bar{\nu}_e$ - spectrum, and in particular,
the appearance of the hard component can be considered as
the signature of
the spin-flavor conversion.

According to fig.~8, the spin-flip  may result in a variety
of distortions of the $\bar{\nu}_e$-spectrum. In particular,
when  the transition probability $P_s$ is constant,
the  permutation of the $\bar{\nu}_e$-
and $\nu_{\mu}$-spectra can be symmetric,
so that $F(\bar{\nu}_e) \simeq (1 - P_s) F_m + P_s F_h$,
Such an effect is realized if the  spectra are in the
asymptotic region
of the suppression pit or
in the region of strong (complete)
transformation.
The
transition can be asymmetric, so that in $\bar{\nu}_e$-spectrum the
suppression of the $F_m$-component is weaker than the appearance of $F_h$-%
component and vice versa. This is realized, when  the spectra are at the
edges of the suppression pit or in the  modulated resonance region.
Note that the typical energy scale of the modulations of the probability
can be characterized by  factor 2 - 3. Therefore the spin-flavor transition
results in a smooth distortion of each component.
The fine structure of the energy spectrum ($\Delta E \sim 1 - 2$ MeV)
can be due to the jumps of density or/and the field twist
in the inner parts of the star (see sect.~7).

Two remarks are in order. The  change of the
$\bar{\nu}_e$-spectrum is expected also if flavor  mixing
is large. In this case, however,  one gets an energy independent interchange
of the spectra with the transition probability $P < 0.5$. If
the neutrino mass hierarchy is inverse,  the properties of the transitions
in the neutrino and antineutrino channels should be interchanged.
In particular, $\bar{\nu}_e$-spectrum will acquire the soft component.

According to (33),  the final $\nu_e$-spectrum is, in general, an
energy dependent combination of all three original spectra.
However, $F_m$ component appears only if there are both  spin-flavor
and flavor transitions. Also the
spin-flavor transition results in appearance of the
$F_m$ component in final $\nu_{\mu}$-spectrum.

The modifications of the spectra
are especially simple,  when
the resonant transitions are either completely
efficient  or completely inefficient: $P_i = 0$ or 1 \cite{AB}.
In this case a complete permutation of
the original spectra occur. Moreover, as follows
from (33, 34), there are
only four possible types of final spectra (Table I).

\begin{table}
\begin{tabular}{|l|l|l|l|l|}
\hline
\sl \# & \sl $F(\nu_e)$ & \sl $F(\bar{\nu}_e)$ & \sl $F(\nu_{ne})$ &
\sl transitions  \\
\hline
1  & $F_s$  & $F_m$   & $4F_h$       & no transitions \\
2  & $F_h$  & $F_m$   & $3F_h + F_s$ & $f;~ f';~ f'f$ \\
3  & $F_s$  & $F_h$   & $3F_h + F_m$ & $s;~ s';~ s's$ \\
4  & $F_m$  & $F_h$   & $3F_h + F_s$ & $sf;~ s'f';~ f'sf;~ s'f's$ \\
5  & $F_h$  & $F_h$   & $2F_h + F_s + F_m$ & $s'f;~ f's;~ s'fs;~
f's'f;~ f's'fs$\\
\hline
\end{tabular}
\caption{
The final neutrino spectra in the case when
transitions are either  complete or completely inefficient:
$P_i \approx 0$ or 1.
In the fifth column we  give a list of the transitions which result
 in a given final spectrum. Here $f$ and $f'$ denote the flavor transitions
$\nu_e - \nu_{\mu}$ and
$\nu_e - \nu_{\tau}$, $s$ and $s'$ denote the spin-flip transitions
$\bar{\nu}_e - \nu_{\mu}$ and
$\bar{\nu}_e - \nu_{\tau}$, respectively;
$sf$ denotes a combination of the two complete transitions:
first $\bar{\nu}_e-\nu_\mu$ and then $\nu_e - \nu_\mu$ etc.
}
\end{table}

According to the Table I, the flavor transitions ($\#$ 2) result in a hard
$\nu_e$-spectrum, whereas $\bar{\nu}_e$-%
spectrum is unchanged. Correspondingly, the $\nu_{ne}$%
-flux acquires a soft component. The neutronization peak consists
of $\nu_{ne}$-neutrinos. The
spin-flavor conversion only ($\#$ 3) leads to a  hard
$\bar{\nu}_e$-spectrum and to unchanged $\nu_e$-spectrum. The flavor
conversion in the inner part of star could be efficient
during  the early stage of the burst, so that
the spectra \#3 for cooling stage  can be accompanied by
the $\nu_{ne}$-neutronization peak.
The  spin-flavor and the subsequent flavor transitions of
$\bar{\nu}_e$ in the inner or  outer parts of the star result in the
$\nu_e$-spectrum coinciding with the initial
$\bar{\nu}_e$-spectrum, $F_m$ ($\#$ 4). In this case
$\bar{\nu}_e$ has hard spectrum $F_h$. The same final spectra ($\#$ 4)
appear also if in addition some
other transitions take place which do not influence the neutrino flux
originally produced as  $\bar{\nu}_e$.
The flavor transition of $\nu_e$
and the spin-flavor transition of $\bar{\nu}_e$
result in the same hard spectra for $\nu_e$ and $\bar{\nu}_e$
($\#$5).
In principle, future
experiments will be able to distinguish these possibilities.

If the transitions are incomplete, the final spectra are certain
energy dependent combinations of the above five spectra.

The spectra can show a strong time dependence. The original
spectra themselves depend on time: the temperatures decrease,
and moreover, they decrease differently for different neutrino species.
Also the transition probabilities  change with time
due to the variations of the effective potential profile in the inner part.

Interesting effect can be related to the existence
of a region in which the density increases with distance. As we
have noticed, here the order of resonances is changed:
neutrinos first cross the flavor
resonance and then the spin-flavor resonance.
In this case $\nu_e$ can be transformed into $\bar{\nu}_e$;
the final $\bar{\nu}_e$-spectrum will contain the soft component,
and moreover the neutronization peak will consist of $\bar{\nu}_e$.
Such a modification of spectra can be realized, if two inner
resonances
$s$ and $f$,  are inefficient. They are at the surface of the protoneutron
star, where the density is changed very quickly,
and therefore the adiabaticity could be  strongly broken.

\section{Effects of magnetic field twist}

\subsection{Field twist. Scale of twist}

If the direction of the magnetic strength lines changes with
distance, the  propagating neutrinos feel the rotation (twist)
of the magnetic field. The field twist leads to the
splitting of the levels with different helicities by the value
$\dot{\phi}$ (see (15)) \cite{APS}.
The effect can be considered as the
modification of the effective potential:
\begin{equation}
 V_{SF} \rightarrow V_{\phi} = V_{SF} + \dot{\phi}.
\end{equation}
Thus the field twist  changes the level crossing  picture
(the positions of resonances).
However, since the flavor and spin-flavor resonances are strongly
separated due to the isotopical neutrality of the medium, the shift (34) hardly
induces the spatial permutation of resonances \cite{APS}.

The field twist can be characterized by the scale of twist, $r_{\phi}$:
\begin{equation}
 r_{\phi} \equiv \frac{\pi}{\dot{\phi}}~,
\end{equation}
so that on the way, $r_{\phi}$, the total rotation angle
(in case of uniform rotation) equals $\Delta \phi  = \pi$.

It is natural to suggest that the
total rotation angle is restricted by
\begin{equation}
 \Delta \phi  \leq  \pi.
\end{equation}
For example, the twist appears, when  the neutrinos cross the toroidal
magnetic  field with  strength lines winding around the torus.
In this case the maximal rotation angle is $\pi$, i.e. the bound (36) is
fulfilled.
Let $\Delta r_{B}$ be the size of the region with the magnetic field,
then $\Delta \phi  \sim  \dot{\phi}\Delta r_{B}$,
and the condition (36) implies
\begin{equation}
r_{\phi} {\buildrel > \over {_{\sim}}}    \Delta r_{B}.
\end{equation}
That is  the scale of the rotation is comparable or larger
than the region with the magnetic field.
For the global field one has  $\Delta r_{B}\, \sim \, r$,
and therefore
\begin{equation}
r_{\phi} \geq  (0.1-1) r.
\end{equation}
As we will see the bound on the total rotation angle (36) restricts strongly
the effects of the field twist.

\subsection{Effect of density suppression}

Let us consider  possible increase of sensitivity to $(\mu B)$
due to the field twist.
According to (34) at $\dot{\phi} = \dot{\phi}_c$, where
\begin{equation}
 \dot{\phi}_c  =  - V_{SF}(r),
\end{equation}
the matter effect is completely compensated at the point $r$.
Let us define the critical rotation scale, $r^{c}_{\phi}$,
at which the condition (39) is fulfilled:
\begin{equation}
 r^{c}_{\phi} =  -\frac{\pi}{V_{SF}}.
\end{equation}
(Note that $r_{\phi}^c$ coincides up to the
factor 2 with the refraction length). At
$r_{\phi} \sim  r^{c}_{\phi}$ the effect of field twist can be
essential, and for
$r_{\phi} < r^{c}_{\phi}$, it even dominates over
the density effect. Comparing $r^{c}_{\phi}$ with
the distance from the center, $r$, we find
(fig.~9) that  $r^{c}_{\phi}/r \geq 1$ for
$r \geq  5\cdot 10^{-2}R_{\odot}$,  and
$r^{c}_{\phi} \ll  r$ for  $r < 5\cdot 10^{-3}R_{\odot}$.
In the inner parts of the star the critical scale is much
smaller than the distance from center, e.g.,
at $r = 10^{-3}R_{\odot}$ one gets  $r_{\phi} \simeq  20$ cm.
The condition (38) is fulfilled for  $r  \simeq  5\cdot
10^{-3}R_{\odot}$. Let us stress that the isotopical
neutrality essentially enlarges the region,  where (38) is
satisfied and therefore the field twist can be important.

Suppose the equality (39) is fulfilled in
some layer of size $\Delta r$
(evidently, the
field twist should be  nonuniform).
In $\Delta r$ the spin-flip effect has a character of precession
with maximal depth, and the sensitivity to $\mu B$ is maximal.
Let us estimate  $\mu B$ taking into account the
restriction (36).  The total rotation angle $\Delta \phi$ equals
\begin{equation}
 \Delta \phi  = \int_{\Delta r} \dot{\phi}(r){\rm d}r =
-  \int_{\Delta r}\,V_{SF}(r){\rm d}r,
\end{equation}
and if $\Delta r \ll r$:
\begin{equation}
 \Delta \phi  \sim  -V_{SF}~\Delta r.
\end{equation}
The spin-flip probability is of the order one, when
$\Delta r \simeq  \pi (2 \mu B)^{-1}$.
Substituting this $\Delta r$ in (42), we
get
the relation:
\begin{equation}
 \mu B \approx \frac{\pi V_{SF}}{2 \Delta \phi}  ~~.
\end{equation}
Finally, (43) and
the bound on the total angle of the twist (36) give
$
\mu B  >  V_{SF}/2
$
which precisely coincides with the  precession bound (6).
Thus, the field twist can relax the precession bound by factor
1/2 at most,  and to the further increase of the sensitivity one should admit
the
field rotation on the total angle which exceeds $\pi$.

In the external layers ($r \geq 0.1R_{\odot}$),  $\Delta r$ is larger than
$r$, and the  compensation (39) implies
fine tuned profiles of $\dot{\phi}(r)$ and
$V(r)$ in a wide spatial region which  seems rather unnatural. On the other
hand, if
the compensation takes place in the inner part ($r <\, 0.1R_{\odot}$)
then $\Delta r\ll r$. However, here
the precession bound is more stringent than the adiabaticity bound,
and the compensation does not allow to gain
in the diminishing of the field strength. In the inner part
the effect of the field twist can be due to influence on the
adiabaticity.

\subsection{Influence of field twist on the adiabaticity}

Nonuniform field rotation $(\ddot{\phi} \neq
0)$ modifies  the  adiabaticity condition. The adiabaticity
parameter equals:
\begin{equation}
 \kappa_{\phi} = \frac{2(2\mu B)^{2}}{|\dot{V}_{SF} + \ddot{\phi}|}.
\end{equation}
At
\begin{equation}
 \ddot{\phi} \simeq -  \dot{V}_{SF}
\end{equation}
$\kappa_{\phi} \rightarrow \infty$, i.e. there is a strong improvement of
the
adiabaticity. The field twist results in flattening of the potential,  so
that $\mbox{d}V_{\phi}/\mbox{d}r  \simeq  0$ or $V_{\phi} \simeq $
const., in some region $\Delta r$.
Let us estimate the minimal value of the total rotation
angle in this region. For this we
suggest that $\dot{\phi} = 0$ at the one of the edges of
$\Delta r$. Then
from  (45), it
follows that $\Delta \phi  \approx  \dot{V}_{SF} (\Delta r)^{2}/2$.
Inserting $\Delta r \approx \pi (2\mu B)^{-1}$ (the condition for the strong
precession effect) in
the last expression,  we get
\begin{equation}
\Delta \phi  \approx \frac{\pi^2 \dot{V}_{SF}}{2(2\mu B)^{2}} \sim
  \frac{\pi}{\kappa_{R}},
\end{equation}
i.e. the total rotation angle is the inverse value of the adiabaticity
parameter without field twist. Therefore,
if the adiabaticity is strongly broken,   $\kappa_{R} \ll 1$,
one needs $\Delta \phi  \gg  \pi$
to get an appreciable spin-flip effect.
For restricted values of total rotation angle
(36) only weakly broken adiabaticity can be restored.
The relation (46) means that if the transition without field twist
is weak ($\kappa \ll 1$), then field twist
will induce a weak effect for $\Delta \phi < \pi$. On the other hand,
if the transition without twist is strong ($\kappa_R \sim 1$), then
the field twist can make it even stronger but the absolute change
of  probability turns out to be always small (fig.~11).

The effect of the adiabaticity restoration due to field
twist can be considered as
the effect of precession in the  region $\Delta r$
with flat potential ($V_{\phi}$ = const.).
For  constant $B$ (which we suggest for simplicity) the
transition   probability equals
\begin{equation}
P_s = \frac{(2\mu B)^{2}}{(2\mu B)^{2}+(V_{\phi}-\frac{\Delta
 m^{2}}{2E})^{2}}\cdot \sin^{2}\sqrt{(2\mu
 B)^{2}+ \left(V_{\phi}-\frac{\Delta m^{2}}{2E}\right)^{2}}
\frac{\Delta r}{2}~~.
\end{equation}
One can find from  (47) that field twist  leads
to the distortion of the energy dependence of the probability in
some  energy interval $\Delta E$ located at
$E/\Delta m^{2} \simeq (2V_{\phi})^{-1}$. Let us estimate the size
of $\Delta E$. According to (47) $\Delta E$
 is  determined by the resonance width $\Delta E/E\, \sim
4 \mu B/V_{SF}$. In turn, $2\mu B$ can be estimated from (46) and
the condition $\Delta \phi < \pi$: $2\mu B {\buildrel > \over {_{\sim}}}
\sqrt{(\pi \dot{V}_{SF})/2}$ .
So that finally, we get
\begin{equation}
\frac{\Delta E}{E} {\buildrel > \over {_{\sim}}}
\frac{\sqrt{2 \pi \dot{V}_{SF}}}{V_{SF}} =
\sqrt{
\frac{2 \pi}{H_{SF} V_{SF}}
}
\end{equation}
If  $V_{SF} \propto r^{-3}$ and $H_{SF} \propto r$, the relation (48)
gives  $\frac{\Delta E}{E} \propto r$. The closer the region with field
twist to the center of the star  the thinner the peak. For $r \sim
10^{-2}R_{\odot}$, one gets
$\frac{\Delta E}{E} {\buildrel > \over {_{\sim}}}    0.3$.
In external region of star $(r > 0.1R_{\odot})$, the field twist
results in
smooth change of the probability in a wide energy region, and  it is
impossible to distinguish it from other  effects (fig. 11).
In any case the absolute value of $\Delta P$ is small, unless
$\Delta \phi \gg \pi$.

As follows from (44) the field twist can also destroy the adiabaticity,
when $|\ddot{\phi}| \gg |\dot{V}|$. Randomly twisting field inhibit the
conversion in such a way. This can result in bumps
of the survival probability that may be similar to those due to density
jumps.

\section{Conclusion}

1. For neutrino mass squared differences
$\Delta m^2 < 10$ eV$^2$
(which are interesting for the cosmology, as well as for the physics of solar
and
atmospheric neutrinos)
the resonant spin-flavor transitions
($\nu_{e}\rightarrow \bar{\nu}_{\mu}$ etc.)
take place in almost isotopically
neutral region of collapsing star.
In this region  which extends from
$\sim 10^{-3} R_{\odot}$ to
$\sim R_{\odot}$
the deviation from the isotopical neutrality, $2Y_{e}-1$,
can be as small as $10^{-3} - 10^{-4}$.  Correspondingly,
the  matter potential  for the spin-flavor
transitions,  being proportional to ($2Y_{e}-1$), turns out to
be suppressed by 3 - 4 orders of magnitude.
Moreover, the potential changes the sign at
the inner edge of the hydrogen envelope.

2. The suppression of the
effective potential in the isotopically neutral region
diminishes the  values of ($\mu$B) needed to induce an
appreciable spin-flip effects by 1.5 - 2 orders of magnitude.
Thus,  the sensitivity of the neutrino burst studies
to the transition magnetic moments of neutrinos increases,
being  of the order $10^{-13}\mu_B$ for
$\Delta m^{2} = (10^{-8} - 10^{-1})$ eV$^2$ and for a reasonable
strength of the magnetic field.
In particular, for $\Delta m^{2} = (10^{-8} - 10^{-9})$ eV$^2$
the desired values of ($\mu$B) turn out to be  $1.5 - 2$ orders of
magnitude smaller than those for a strong conversion of the solar
neutrinos.

3. In the isotopically neutral region the
potential changes very quickly in the layers with local
ignition (jumps of the potential).
The jumps  result in distortion of the energy dependence
of probabilities. In particular,  one can expect the
appearance of  thin peaks in the survival probability if the
jump is situated in the inner part of the star.

4. Depending on the values of the neutrino parameters as well as  on the
magnetic field profile one expects a variety of the modifications
of the neutrino spectra.
In the case of a direct mass hierarchy
and a small flavor mixing the main
signature of the spin-flip effect is the distortion of the
$\bar{\nu}_e$-energy spectrum, and especially the appearance
of the high energy tail.
In general, the final $\bar{\nu}_e$%
-spectrum is the energy dependent combination  of the
original $\bar{\nu}_e$-spectrum and the hard spectrum of the
non-electron neutrinos.
Another important signature of the spin - flip can be obtained from the
comparison of the spectra of different neutrino species.
In particular, the $\bar{\nu}_e$-  and $\nu_{\mu}$-%
spectra can be completely permuted.
The combination of the spin-flip effect with
other (flavor) transitions  may result in rather peculiar final
spectra. For example,
${\nu}_e$ may have the spectrum of the original
$\bar{\nu}_e$, whereas $\bar{\nu}_e$ may have the original $\nu_{\mu}$-%
spectrum.
The electron neutrino and antineutrino spectra can be the
same and coincide with the hard spectrum of the original muon neutrinos,
etc.

5. The resonant spin-flip effect for the massless or the degenerate ($\Delta
m^{2} {\buildrel < \over {_{\sim}}}
10^{-10}$ eV$^{2}$) neutrinos can be induced by the change of
the nuclear
composition. Such a transition takes place both for neutrinos and
antineutrinos in the same spatial region (near to the bottom of the
H-envelope),  resulting again in hard and equal
${\nu}_e$- and $\bar{\nu}_e$-spectra.

6. The absence of the considered effects allows one  to get the
bound on ($\mu B(r)$) as a function of $\Delta m^{2}$. We
estimate such a bound for SN $1987$A.

7. In the isotopically neutral region, where the
matter potential is strongly suppressed, the effects of (even) small
field twist may be important.
Field twist can further suppress the potential, thus increasing
the sensitivity to the magnetic moment. However,  if the total rotation
angle is restricted ($<\pi$), a possible diminishing of
$\mu B$ can be by factor 1/2  at most.
The twist of the field may induce a distortion
of the neutrino energy spectra. In particular,
improving  adiabaticity,
in the inner part of the isotopically neutral region the field twist
may lead  to peaks of the transition probabilities.

\section*{Acknowledgements}
The authors are grateful to E. Kh. Akhmedov for discussions
and for numerous remarks concerning this paper.
H. A.  would like to thank Prof. A. Salam, the International
Atomic Energy Agency and UNESCO for hospitality at the International
Centre for Theoretical Physics.\\

\pagebreak

{\large {\bf Figure Captions}}

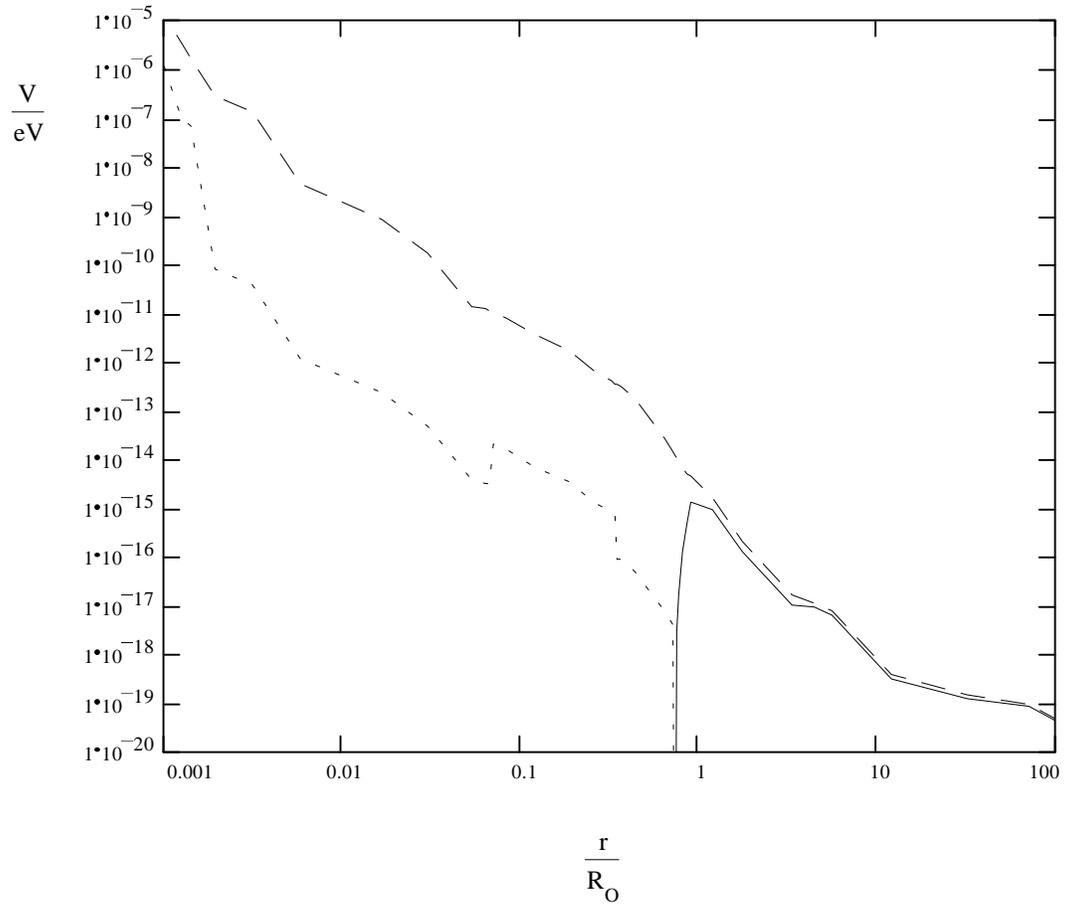
Figure 1.
The dependence of the potentials for the spin-flavor conversion,
$\bar{\nu}_{e}\rightarrow {\nu}_{\mu}$,
(dotted and solid lines) and the flavor conversions, $\nu_{e}\rightarrow
\nu_{\mu}$, (dashed line) on the distance from the
center of the star $(M = 15M_{\odot})$.
At $r \simeq 0.8 R_{\odot}$,
the spin-flavor potential changes the sign, and for $r > 0.8R_{\odot}$
we depict $- V_{SF}$ which coincides with $V_{SF}$ for
$\nu_{e}\rightarrow \bar{\nu}_{\mu}$ channel.

Figure 2.
The qualitative dependence of the energy levels
on the distance from the center of the star, $r$,  for different
values of
$\Delta m^{2}/2E$. The energies are defined  in such a way that
$H(\nu_{e})(r) = 0$  (solid line).
(The energy scale is relative).
Dotted lines show a behaviour of the eigenvalues of the Hamiltonian
near the crossing points. Also marked are the positions of the inner and
outer edges  of the isotopically neutral region.
a). $E/\Delta m^{2} \approx 0$; b).
$E/\Delta m^{2} {\buildrel < \over {_{\sim}}}
 3\cdot10^{-8}$ eV$^2$;
c). $E/\Delta m^{2} {\buildrel > \over {_{\sim}}}
 3\cdot 10^{-8}$ eV$^2$;

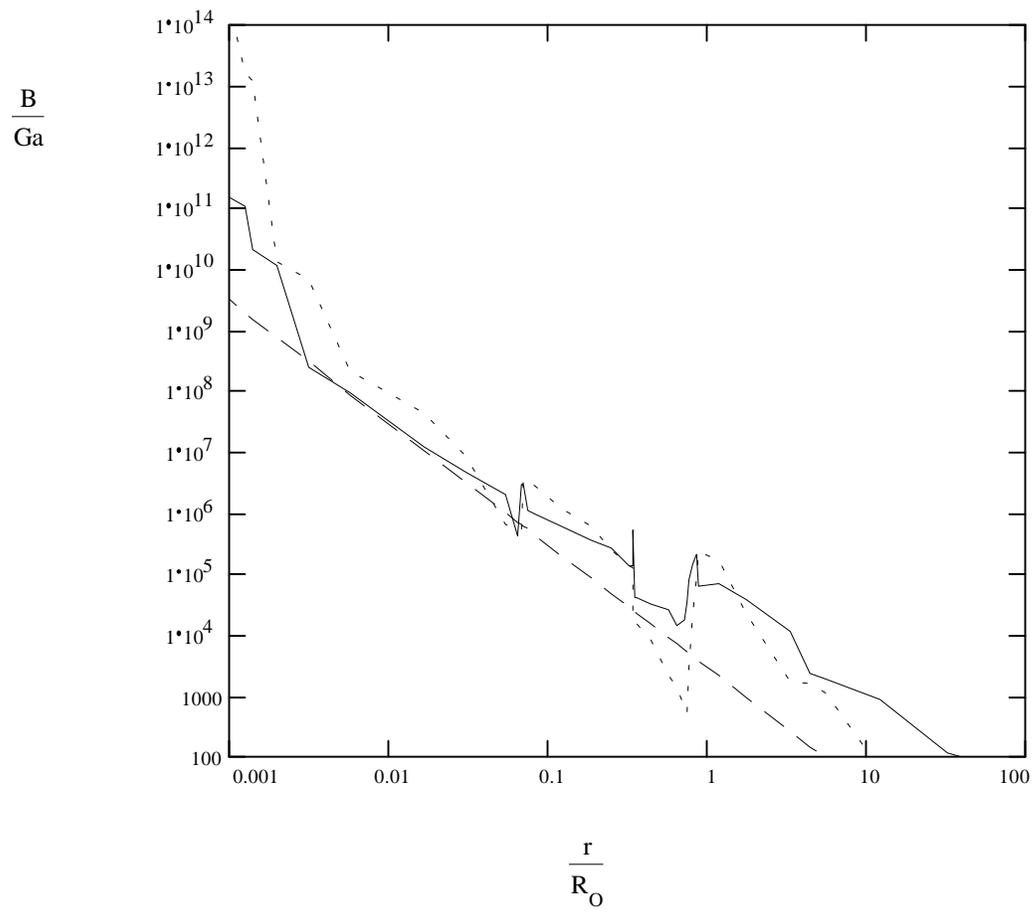
Figure 3.
The dependence of the adiabaticity $B_A$ (solid line) and the
precession $B_P$ (dotted line) bounds for
$\mu_{\nu} = 10^{-12} \mu_{B}$
on the distance from the
center of  star.
Also shown is the magnetic field
profile  (2) with $B_{0} = 1.5\cdot 10^{13}$ Gauss and $k = 2$ (dashed
line).

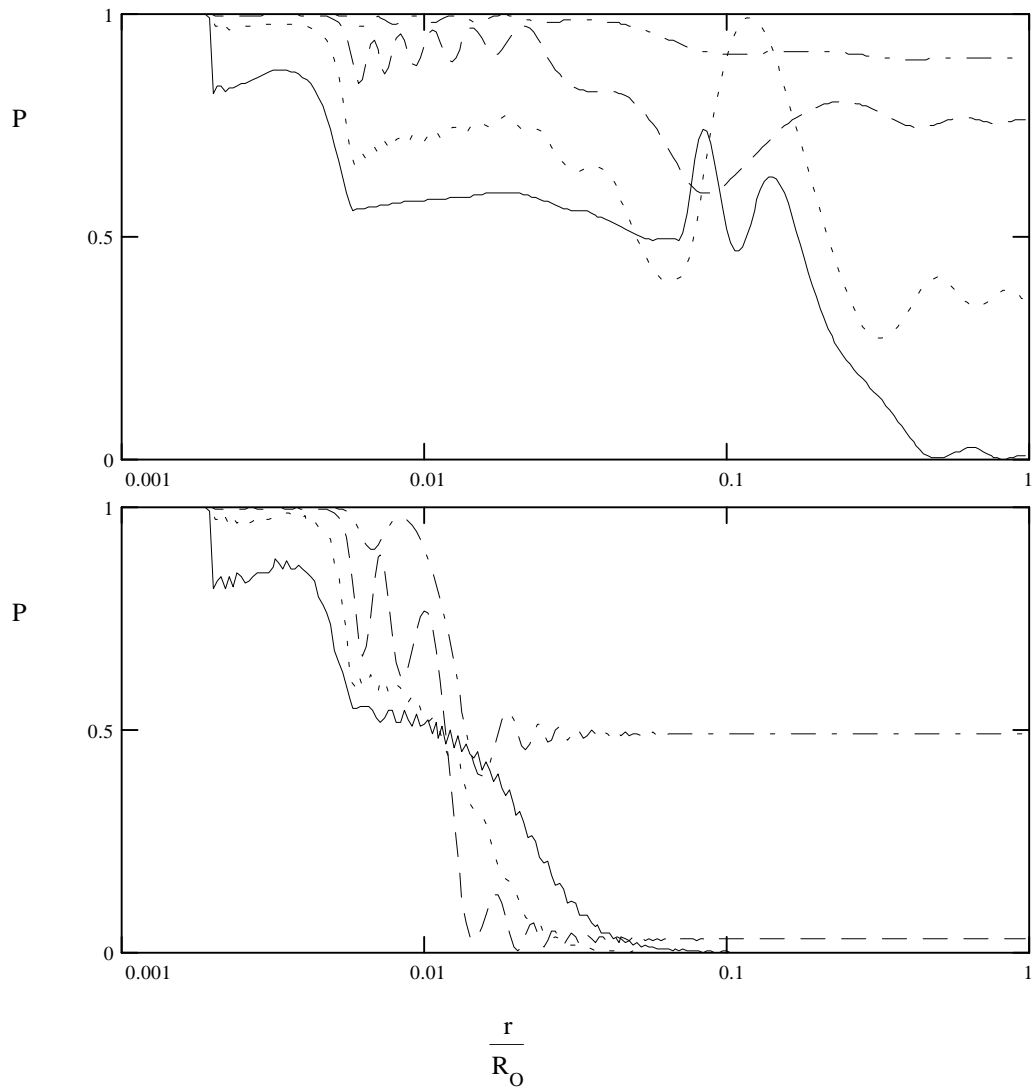
Figure 4.
Spatial picture of the ($\bar{\nu}_{e}\rightarrow \nu_{\mu}$)%
-spin-flavour transition. The dependence of the
$\bar{\nu}_e$-survival probability on the distance from the center
of the star for different strengths of the magnetic field. The field profile
(2) was used with $k = 2$ and
$B_0$ (in Gauss):
$5\cdot10^{12}$ (dashed-dotted line), $1.5\cdot10^{13}$ (dashed line),
$5 \cdot 10^{13}$ (dotted line), $1.5 \cdot 10^{14}$ (solid line).
a). $E/\Delta m^2 = 10^8$ MeV/eV$^2$
b). $E/\Delta m^2 = 10^6$ MeV/eV$^2$.

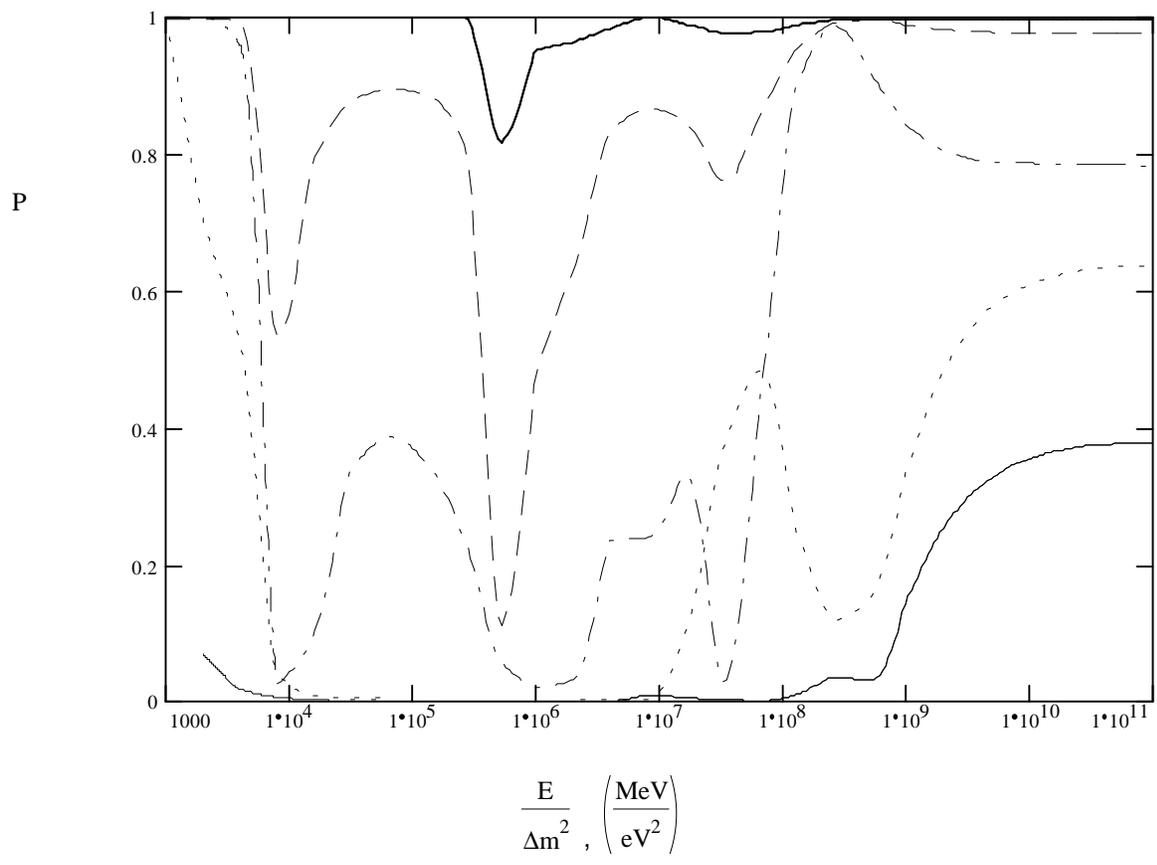
Figure 5.
The $\bar{\nu}_{e}$-survival probability for the
($\bar{\nu}_{e}\rightarrow \nu_{\mu}$)-%
spin-flip transition in the global magnetic field
as a function of $E/\Delta m^{2}$ for different strengths
of the magnetic field.
The magnetic field profile (2) was used with $k = 2$ and
$B_0$ (in Gauss):
$1.5 \cdot 10^{12}$ (bold solid line),
$5 \cdot 10^{12}$ (dashed line), $1.5 \cdot 10^{13}$ (dashed-dotted line),
$5 \cdot 10^{13}$ (dotted line), $1.5 \cdot 10^{14}$ (solid line).

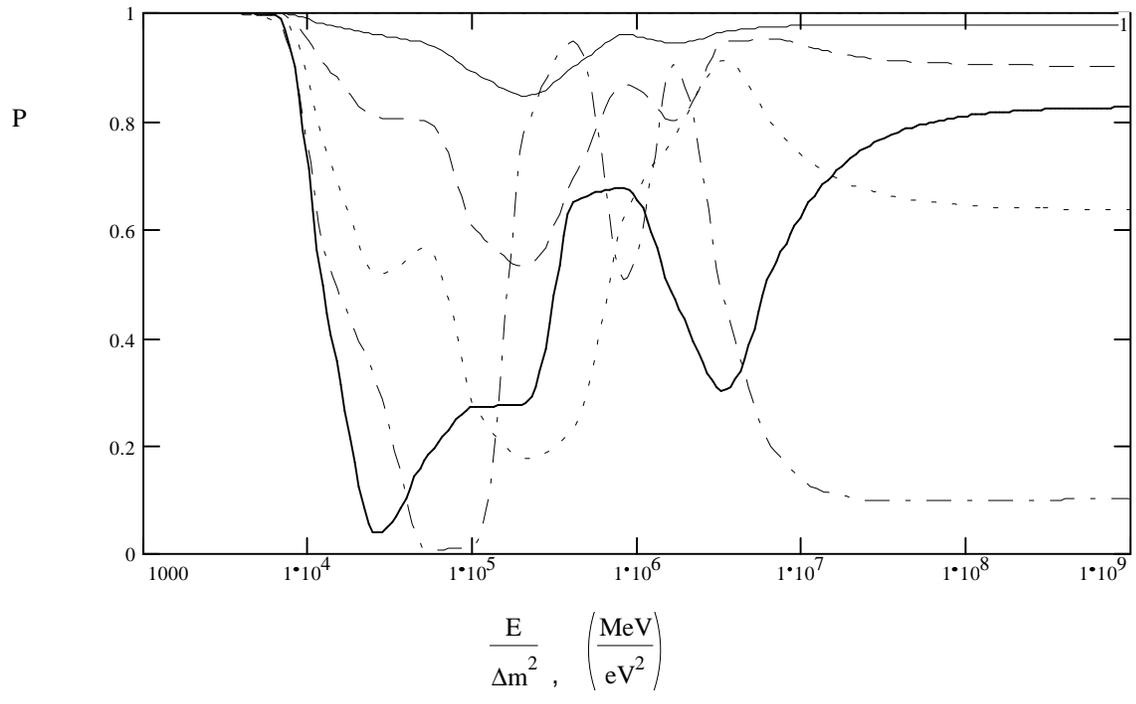
Figure 6.
The $\bar{\nu}_{e}$-%
survival probability for the
($\bar{\nu}_{e}\rightarrow \nu_{\mu}$) - spin-flavor transition,
in the local magnetic field
as functions of $E/\Delta m^{2}$
for different values of the magnetic field.
A constant magnetic field $B_c$ in the internal
($1 - 3)\cdot 10^{-2}R_{\odot}$ was used. $B_c$ (in Gauss):
$3 \cdot 10^{7}$ (solid line),
$6 \cdot 10^{7}$ (dashed line), $2 \cdot 10^{8}$ (bold solid line),
$3 \cdot 10^{8}$ (dashed-dotted line).

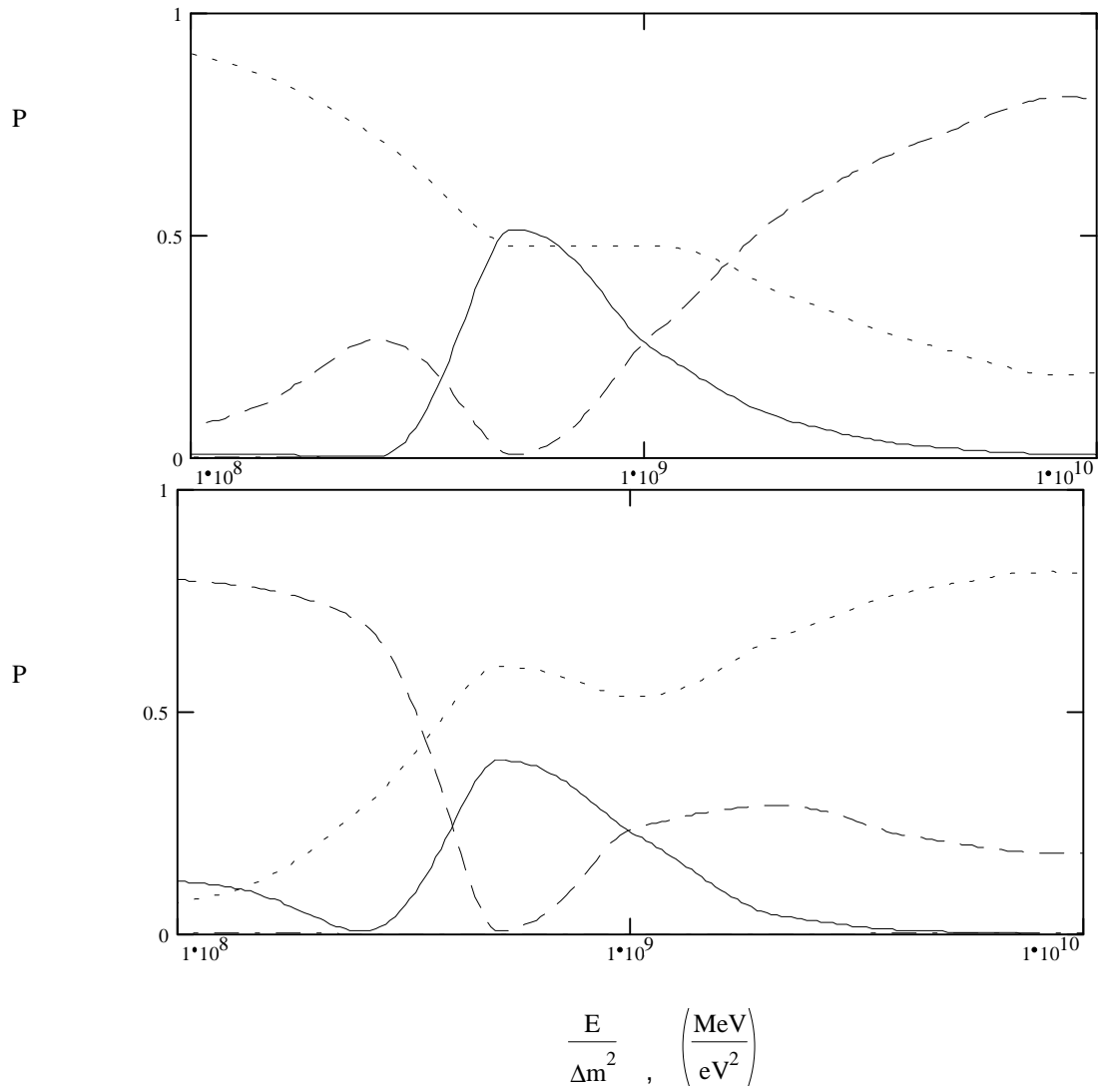
Figure 7.
The probabilities of different transitions for  the neutrino system with
a magnetic moment and a vacuum mixing as functions of
$E/\Delta m^2$. The initial state is (a)
$\bar{\nu}_{e}$, and (b)  $\nu_{\mu}$.
The lines  show the probabilities to find in the final state
$\nu_e$ (solid),  $\bar{\nu}_{e}$ (dotted),
and $\nu_{\mu}$ (dashed line).
A vacuum mixing angle  $\sin^{2}2\theta  = 10^{-2}$
and a constant magnetic field $B_c = 10^{4}$ Gauss in the interval
$R =  (0.1-1.3)R_{\odot}$ were used.

\clearpage

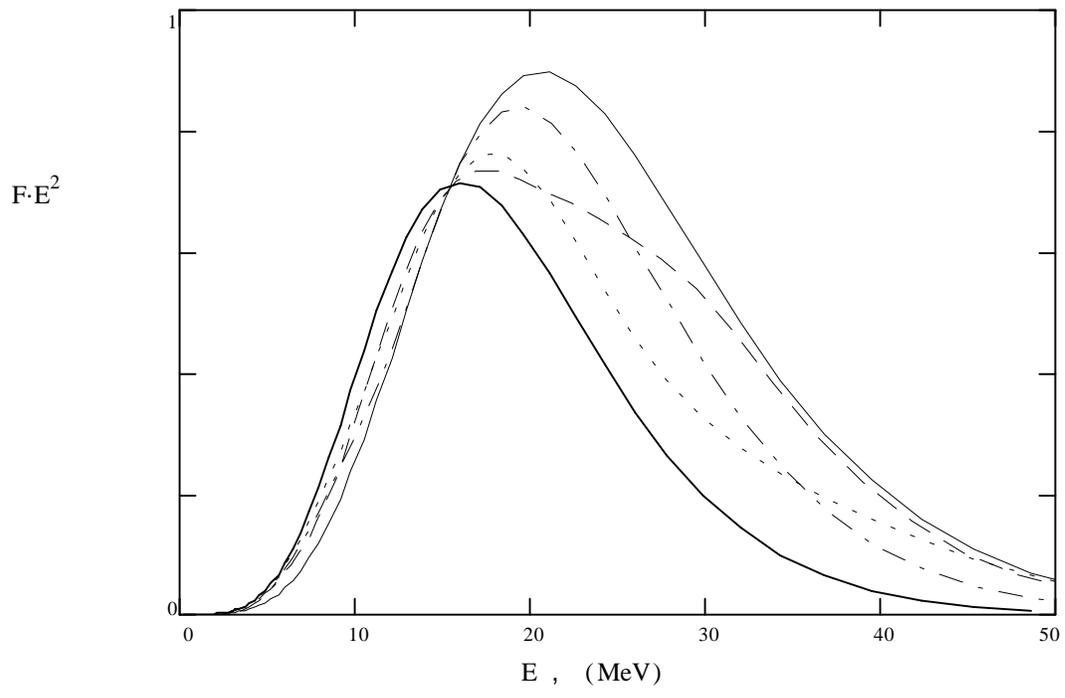
Figure 8.
The distortion of the
$\bar{\nu}_e$%
-spectrum.
The dependence of the product (flux)$\times$(energy squared)
on energy for different sets of  neutrino parameters and
different  configurations of the magnetic fields. The original
$\bar{\nu}_e$-spectrum is shown by the bold solid line.
The case of the complete $\bar{\nu}_e - \nu_{\mu}$ transformation
($F = F_h$) is shown by the solid line.

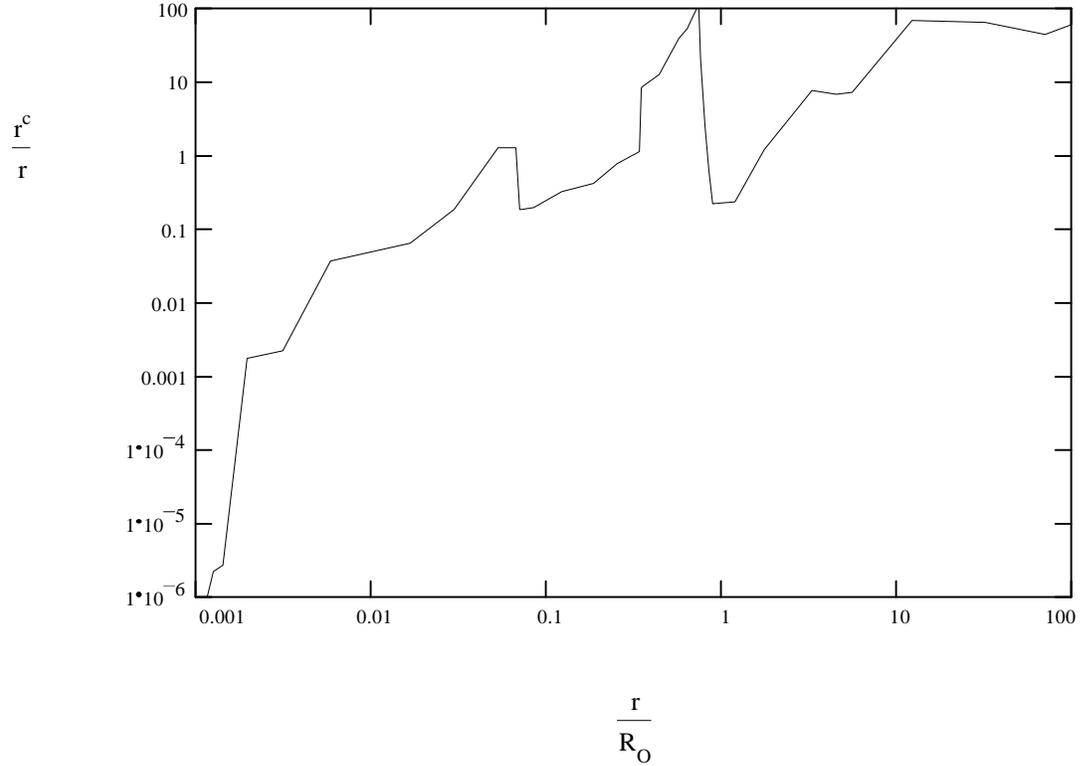
Figure 9.
The critical scale of the field twist over the
   distance from the center of the star  as a  function of  $r$.

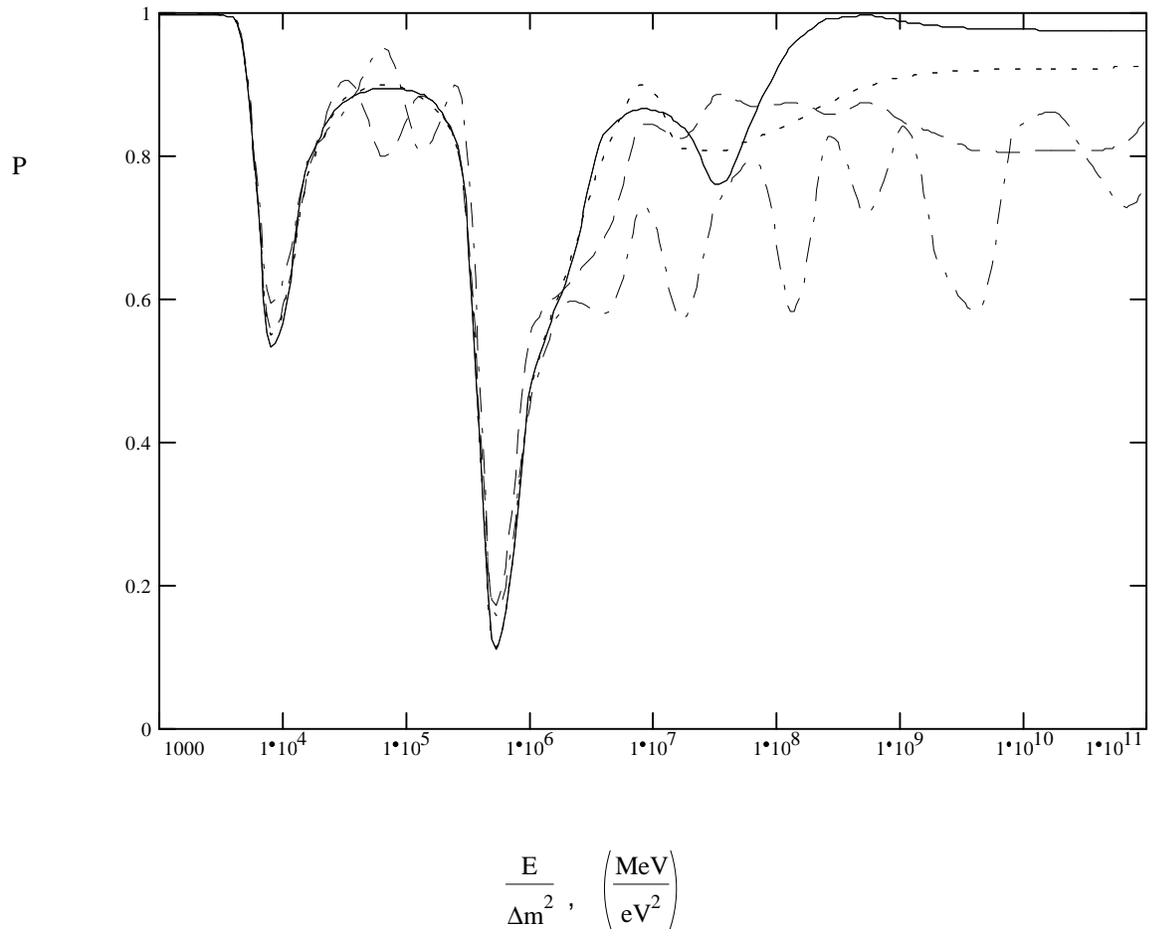
Figure 10.
The effects of the non-uniform field twist. The $\bar{\nu}_{e}$-%
survival probability of the spin-flip transition,
$\bar{\nu}_{e}\rightarrow \nu_{\mu}$ ,
as a function of $E/\Delta m^{2}$ for different
configurations (scales) of the field twist.
The field profile (2) with
$k = 2$ and $B_0 = 5\cdot 10^{12}$ Gauss was used. The solid line shows
the probability without field twist. The field twist is located in
the region $r_0 - (r_0 + \delta r)$, and the  profile of twist
is described by $\dot{\phi}(r) = (2\pi/\delta r) (1 - (r-r_0)/\delta r)$;
the total rotation angle equals $\pi$.
The curves correspond to
$r_0 = 5 \cdot 10^{-3} R_{\odot}$ and  $\delta r/ R_{\odot}$ = $10^{-4}$
(dotted), $5 \cdot 10^{-4}$ (dashed), $2.5 \cdot 10^{-3}$ (dashed-dotted).


\begin{thebibliography}{99}
\bibitem{van} S. Van den Bergh, Phys. Rep. {\bf 204}, 385 (1991).
\bibitem{kam} K. Hirata et al., Phys. Rev. Lett. {\bf 58}, 1490 (1987).
\bibitem{imb} R. M. Bionta et al., Phys. Rev. Lett. {\bf 58}, 1494 (1987).
\bibitem{baks} E. N. Alexeyev et al.,  Phys. Lett. B {\bf 205}, 209 (1988).
\bibitem{lsd} M. Aglietta et al., Europhys. Lett. {\bf 3}, 1315 (1987).
\bibitem{lvd} C. Bari et al., Nucl. Inst. Meth. A {\bf 277}, 11 (1989);
M. Aglietta et al., Nuov. Cim. A, {\bf 105} 1793 (1992).
\bibitem{sno} Scientific and Technical Description of Mark II
SNO Detector, ed. by E. W. Beier and D. Sinclair,
Report SNO-89-15 (1989).
\bibitem{skam} K. Nakamura, ICRR-Report-309-94-4.
\bibitem{icar} P. Benetti et al., Nucl. Inst. \& Meth. A315, 223 (1992),
ibidem A332, 395 (1993).
\bibitem{DBC}D. B. Cline, preprint UCLA-CAA 107-3/94.
\bibitem{AC}A. Cisneros, Astrophys. Space Sci. {\bf 10}, 87 (1970).
\bibitem{KF}K. Fujikawa and R. E. Shrock, Phys. Rev. Lett. {\bf 45}, 963
(1980).
\bibitem{MBV} M. B. Voloshin, M. I.
Vysotsky, and L. B. Okun, Zh. Eksp. Teor. Fiz. {\bf 91}, 754 (1986) [Sov.
Phys. JETP {\bf 64}, 446 (1986)].
\bibitem{JS}J. Schechter and J. W. F. Valle, Phys. Rev. D {\bf 24}, 1883
(1981); ibidem {\bf 25}, 283 (E) (1982).
\bibitem{lim} C.-S. Lim and W. J. Marciano, Phys.
Rev. D {\bf 37}, 1368 (1988).
\bibitem{Ekh}E. Kh. Akhmedov, Sov. J.
Nucl. Phys. {\bf 48}, 382 (1988); Phys. Lett. B {\bf 213}, 64 (1988).
\bibitem{AB}E. Kh. Akhmedov and Z. G. Berezhiani, Nucl. Phys. B {\bf
373}, 479 (1992).
\bibitem{vol} M. B. Voloshin, Phys. Lett. B {\bf 20}, 360 (1988).
\bibitem{pelt} J. T. Peltoniemi, Astron. \& Astrophys., {\bf 254}
121 (1992).
\bibitem{akh} E. Kh. Akhmedov, Sov. Phys. JETP {\bf 68}, 676 (1989).
\bibitem{mom} M. Fukugita and S. Yazaki Phys. Rev. D {\bf 36}, 3817 (1987);
G. G. Raffelt, Astrophys. J. {\bf 365}, 559 (1990);
V. Castellani and S. Degl'Innocenti, Astrophys. J. {\bf 402}, 574 (1993).
\bibitem{M} M. Moretti, Phys. Lett. B {\bf 293}, 378 (1992).
\bibitem{ST}S. E. Woosley and T. A. Weaver, Annu. Rev. Astron. Astrophys.
{\bf 24}, 205 (1986).
\bibitem{SNT}S. E. Woosley, N. Langer and T. A. Weaver, Ap.\ J. {\bf
411}, 823 (1993).
\bibitem{KS}P. I. Krastev and A. Yu. Smirnov, Mod. Phys. Lett. A {\bf 6},
1001 (1991).
\bibitem{MS}S. P. Mikheyev and A. Yu. Smirnov,
Sov. J. Nucl. Phys. {\bf 42}, 913 (1985).
\bibitem{kras}P. I. Krastev and A. Yu. Smirnov,
Phys. Lett. B {\bf 338} 282 (1994).
\bibitem{atm} R. Becker-Szendy et al., Phys. Rev. D {\bf 46}, 3720 (1992);
Y. Fukuda et al., Phys. Lett. B {\bf 335}, 237 (1994).
\bibitem{mfs} see for latest discussion
E. Kh. Akhmedov, A. Lanza and S. T. Petcov, SISSA Report 169/94/A-EP.
\bibitem{APS} A. Yu. Smirnov, Phys. Lett. B {\bf 260}, 161 (1991);
E. Kh. Akhmedov, S. T. Petcov, and A. Yu. Smirnov, Phys.
Rev. D {\bf 48}, 2167 (1993).
\bibitem{burr} A. Burrows, D. Klein and R. Gandhi, Phys. Rev. D {\bf 45},
3361 (1992).
\bibitem{SSB}A. Yu. Smirnov, D. N. Spergel, and J. N. Bachall, Phys. Rev.
D {\bf 49}, 1389 (1994).
\bibitem{Kh}E. Kh. Akhmedov, Phys. Lett. B {\bf 213}, 64 (1988).
\bibitem{pant} L. B. Okun, Sov. J. Nucl. Phys., {\bf 48} 967
(1990); J. Pantaleone, Phys. Lett. B {\bf 287}, 128 (1992);
S. Samuel, Phys. Rev. D {\bf 48}, 1462 (1993).
\end{thebibliography}
\end{document}